\documentclass[twocol]{ametsocV5}

\usepackage{amsmath,amsfonts,amssymb,bm}
\usepackage{mathptmx}
\usepackage{newtxtext}
\usepackage{newtxmath}

\usepackage{booktabs}
\newcommand{\doublerule}{\toprule\specialrule{\heavyrulewidth}{\doublerulesep}{0.95em}}

\let\oldunderscore\_
\renewcommand\_{\allowbreak\oldunderscore}
\def\param#1{\footnotesize\texttt{#1}}

\newcommand{\secref}[1]{Section~\ref{#1}}
\newcommand{\subsecref}[2]{Section~\ref{#1}\ref{#2}}
\newcommand{\subsubsecref}[3]{Section~\ref{#1}\ref{#2}.\ref{#3}}
\newcommand{\figref}[1]{Fig.~\ref{#1}}
\newcommand{\tabref}[1]{Table~\ref{#1}}

\newcommand{\quantitytable}{Table~A1}

\usepackage[binary-units=true]{siunitx}
\newcommand{\abni}{\ensuremath{\mathrm{AI}}}

\usepackage{verbatim}

\title{FOWD: A Free Ocean Wave Dataset for Data Mining and Machine Learning}

\authors{Dion Häfner\correspondingauthor{Dion Häfner, dion.haefner@nbi.ku.dk}}

\affiliation{Niels Bohr Institute, University of Copenhagen, Copenhagen, Denmark}

\extraauthor{Johannes Gemmrich}
\extraaffil{University of Victoria, Victoria, British Columbia, Canada}

\extraauthor{Markus Jochum}
\extraaffil{Niels Bohr Institute, University of Copenhagen, Copenhagen, Denmark}

\abstract{
The occurrence of extreme (rogue) waves in the ocean is for the most part still shrouded in mystery, as the rare nature of these events makes them difficult to analyze with traditional methods. Modern data mining and machine learning methods provide a promising way out, but they typically rely on the availability of massive amounts of well-cleaned data.
\\ \indent
To facilitate the application of such data-hungry methods to surface ocean waves, we developed FOWD, a freely available wave dataset and processing framework. FOWD describes the conversion of raw observations into a catalogue that maps characteristic sea state parameters to observed wave quantities. Specifically, we employ a running window approach that respects the non-stationary nature of the oceans, and extensive quality control to reduce bias in the resulting dataset.
\\ \indent
We also supply a reference Python implementation of the FOWD processing toolkit, which we use to process the entire CDIP buoy data catalogue containing over 4 billion waves. In a first experiment, we find that, when the full elevation time series is available, surface elevation kurtosis and maximum wave height are the strongest univariate predictors for rogue wave activity. When just a spectrum is given, crest-trough correlation, spectral bandwidth, and mean period fill this role.
}

\begin{document}

\maketitle

%
%
\statement
%

Rogue waves are ocean waves at least twice as high as the surrounding waves. They tend to strike without warning, often damaging ocean-going vessels and offshore structures. Because of their inherent randomness and rarity, there is no satisfying forecasting method for rogue wave risk, nor do we know under which conditions they preferably occur.

Modern machine learning methods provide a promising new alternative, but they require vast amounts of clean data. Here, we provide a way to create such a dataset from ocean surface measurements.

We demonstrate our method by processing a buoy dataset containing over 4 billion wave measurements. The result is freely available for download. In a first experiment, we show that it \emph{is} possible to extract risk factors for rogue waves from data, with some conditions producing 10--100$\times$ more rogue waves than others.

This paves the way to a better physical understanding and forecasting methods of these dangerous events.


%








\section{Introduction}

During the last 25 years, the study of extreme ocean waves (also known as ``rogue waves'' or ``freak waves'') has experienced a Renaissance, triggered by the observation of the \SI{25.6}{\metre} high New Year Wave at the Draupner oil rig in 1995 \citep{haver2004possible}. By now, there are several known mechanisms to generate much higher waves than predicted by linear theory \citep{adcock_physics_2014,kharif_physical_2003,slunyaev_rogue_2011,dysthe_oceanic_2008}, most of which rely on either highly nonlinear effects like Benjamin-Feir instability \citep[e.g.][]{gramstad_modulational_2018} or weakly nonlinear corrections to the Rayleigh wave height distribution \citep[e.g.][]{toffoli_evolution_2010}.

However, while there is plenty of experimental evidence for these mechanisms in wave tanks and simulations, the relative importance of these processes in the real ocean is still unknown. This is evidenced by the rich spectrum of studies emphasizing different physical causes of rogue waves \citep{janssen_extension_2009,toffoli_evolution_2010,gemmrich_dynamical_2011,xiao_rogue_2013,fedele_real_2016,gramstad_modulational_2018,mcallister_laboratory_2019}. This has the consequence that, so far, there is no reliable forecast for rogue wave risk \citep[see also][]{dudley_rogue_2019}, although there have been some recent efforts \citep{barbariol_maximum_2019}.

There are several studies that aim to relate sea state parameters to rogue wave occurrence \citep{cattrell_can_2018,casasprat_short-term_2010,karmpadakis_assessment_2020,gemmrich_dynamical_2011}, but they are limited by the analyzed amount of data (often only one or several storms), their coverage of parameter space (often only look at 1 or 2 parameters), or sophistication of analysis (often no uncertainty analysis). To our knowledge, no study has been able to show the dependence of rogue wave occurrence on sea state (or show that it does not exist) with statistical significance throughout a wide regime of sea states.

We credit this shortcoming to a lack of sufficient amounts of well-curated, accessible data on one hand, and a lack of a sophisticated analysis framework that handles nonlinearities and feature interactions on the other hand. In this study, we address the first issue and present FOWD, a Free Ocean Wave Dataset.

Particularly since the advent of machine learning competitions --- e.g.\ via the platform ``Kaggle'' (\url{kaggle.com}), where teams compete to find the best performing machine learning solutions to domain-specific problems --- freely available, high-quality data sets have become an invaluable resource both as benchmarks for machine learning researchers and as study objects for domain experts. Enabling easy access to domain-specific data allows even non-domain experts to participate in model building, to the benefit of the whole research community. We therefore also see this work as an important stepping stone towards opening extreme wave research to a wider, potentially more machine learning-literate, audience.

Note that, while we will be using rogue waves as a motivating example throughout this publication, other researchers can and should of course use FOWD to study phenomena other than extreme wave / crest heights (e.g., wave steepness or characteristic shape). In essence, FOWD relates aggregated sea state parameters to individual wave measurements. Applications are therefore plentiful.

As a primary data source for this version of FOWD, we will use the CDIP buoy data catalogue. CDIP (Coastal Data Information Program) is a buoy network consisting primarily of Datawell Directional Waverider buoys for wave monitoring around the coasts of the United States \citep[see e.g.][]{behrens_cdip_2019}, maintained by the Scripps Institution of Oceanography. The CDIP catalogue (as of November 2020) contains measurements at 161 locations along the West and East Coasts of North America and US overseas states and territories like Hawaii, Guam, Puerto Rico, and the Marshall Islands.

\secref{sec:specification} describes FOWD in detail, particularly which parameters are included, how they are computed, and which quality control processes we employ to validate the results. \secref{sec:implementation} outlines our Python reference implementation that allows us to efficiently process massive amounts of raw data, and \secref{sec:cdip} describes the processing of the CDIP buoy data catalogue. \secref{sec:application} finally gives an example application where we look at how rogue wave probabilities vary depending on various sea state parameters. \secref{sec:conclusion} gives a summary and conclusive remarks.

\section{The FOWD Specification} \label{sec:specification}

At its core, FOWD describes a mechanism to process raw observations (elevation time series and, optionally, directional spectra) into a catalogue that maps parameters describing the current sea state $x$ to observed wave or crest parameters $y$.

By ``wave'' we denote the series of surface elevations (relative to the \SI{30}{\minute} mean elevation) from a given zero upcrossing to the next zero upcrossing. The crest (trough) is then the maximum (minimum) elevation of the wave, and the wave height is the sum of its crest height and trough depth. Some waves might be excluded due to quality control criteria, see \subsecref{sec:specification}{sec:qc}.

Throughout this study, we characterize extreme waves based on their abnormality index $\abni = H/H_S$, with wave height $H$ and spectral significant wave height $H_S = 4 \sqrt{m_0}$, where $m_0$ is the zeroth moment of the spectral density (see also \subsubsecref{sec:specification}{sec:quantities}{sec:spectral-density}).

FOWD output files are in netCDF4 format, which is widely used throughout the sciences and allows additional metadata to be attached. Every row in the resulting netCDF4 file represents a single wave and the sea state in which it was recorded.

\subsecref{sec:specification}{sec:quantities} introduces the various quantities included in FOWD output, and gives a more in-depth description of the computation of some parameters (where estimation is non-obvious or ambiguous). \subsecref{sec:specification}{sec:window} describes the running window processing approach we use in FOWD. \subsecref{sec:specification}{sec:qc} lists our quality control (QC) criteria, and \subsecref{sec:specification}{sec:reproducibility} outlines the steps we take to ensure reproducibility of FOWD output files.

\subsection{Computed Quantities} \label{sec:quantities}

We group all output quantities into 4 categories:

\begin{enumerate}
    \itemsep1em
    \item \textbf{Station metadata}. Anything that is specific to the sensor (and is not directly related to waves or the sea state). This includes both metadata describing the raw data source (to ensure reproducibility, more in \subsecref{sec:specification}{sec:reproducibility}) and the conditions in which it was recorded (latitude / longitude, water depth).

    \item \textbf{Wave-specific parameters}. These are all quantities that describe a single wave, such as wave height or maximum slope. A typical study using FOWD aims to determine how a wave-specific parameter depends on one or several sea state parameters.

    \item \textbf{Aggregated sea state parameters}. These describe the circumstances in which each wave occurred, i.e., they relate to the past sea state of each wave. They are computed based on the immediate 10 and 30 minute history prior to (but not including) the current wave (see also \subsecref{sec:specification}{sec:window} for more on this running-window approach). Quantities are computed using only the raw sea surface elevation as input (either directly or by computing a spectrum first).

    \item \textbf{Directional sea state parameters}. Some sensors (like the CDIP buoys) might include additional directional information that is not computable from the raw surface elevation time series. When such directional information (in form of a directional spectrum) is given, FOWD computes some directional parameters from it and includes them in the output. Note that this does \emph{not} use the same running window approach as the aggregated sea state parameters. Instead, each wave is mapped to the nearest (in time) available directional measurement. I.e., directional information usually includes some information relating to the \emph{future} of the wave. But since directional information is robust to the influence of individual extreme events, we do not consider this a problem.
\end{enumerate}

A complete overview of all computed quantities is shown in \quantitytable. Here, we outline some important quantities (as suggested in literature) and how they are estimated from the observed time series.

\subsubsection{Space-time domain transformations}

Since FOWD only processes (one-dimensional) point measurements, we need some mechanism to transform information from the time domain back to the spatial domain. We relate frequencies $f$ to wave numbers $k$ (and by extension, periods to wavelengths) through the dispersion relation for linear waves:

\begin{equation}
    f^2 = \frac{g k}{(2 \pi)^2} \tanh(k D) \label{eq:dispersion}
\end{equation}

with water depth $D$ and gravitational acceleration $g = \SI{9.81}{\metre\per\second\squared}$. This also assumes the absence of currents.

To determine the wave number for a given frequency, we use an approximate inverse of \eqref{eq:dispersion} as given in \citet{fenton_numerical_1988}:

\begin{equation}
    k \approx \frac{\alpha + \beta^2 \cdot \cosh^{-2}\beta}{D \left( \tanh\beta + \beta \cdot \cosh^{-2}\beta \right)}
\end{equation}

with

\begin{align}
    \alpha &= (2 \pi f)^2 \frac{D}{g} \\
    \beta &= \frac{\alpha}{\sqrt{\tanh\alpha}}
\end{align}

\subsubsection{Spectral density estimation} \label{sec:spectral-density}

In order to compute spectral quantities, we need to estimate the spectral density $\mathcal{S}(f)$ from the raw surface elevation time series. There is no unique way to do this, and any given method is a trade-off between spectral resolution, bias, and variance (noise).

In FOWD, we chose to use Welch's method \citep{welch_use_1967} with a window length of \SI{180}{\sec} and a window overlap of \SI{50}{\percent} using a Hann, or Hanning, window. This corresponds to about 230 measurements per segment in the case of CDIP data with sampling frequency \SI{1.28}{\hertz}. This implies that the \SI{30}{\minute} spectra are an average of 20 individual segments, and the \SI{10}{\minute} spectra are an average of 7 segments. All segments are zero-padded to the next highest power of 2.

This gives a spectral resolution of \SI{0.005}{\hertz} and a maximum (Nyquist) frequency of \SI{0.64}{\hertz} for \SI{1.28}{\hertz} CDIP data.

We can then compute moments of $\mathcal{S}$ by integrating:

\begin{equation}
    m_n = \int_0^\infty f^n \mathcal{S}(f) \;\mathrm{d}f
\end{equation}

We numerically approximate all integrals in FOWD through a trapezoidal rule (with second-order accuracy).

\subsubsection{Wave Period and Steepness}

There are several popular approaches to define a dominant wave period for a given sea state. Depending on the application, either peak period, spectral mean period, or mean zero-crossing period may be more appropriate. Also, since we only have access to a noisy estimate of the true spectral density $\mathcal{S}$, some ways to compute the mean period from the spectrum are more accurate than others, depending e.g.\ on the frequency resolution of the sensor.

Therefore, we include several estimates of dominant wave period / frequency in FOWD:

\begin{align}
    \text{Spectral peak period} \qquad & \overline{T}_p = \frac{\int_0^\infty \mathcal{S}(f)^4 \;\mathrm{d}f}{\int_0^\infty f \mathcal{S}(f)^4 \;\mathrm{d}f} \label{eq:peak-period} \\
    \text{\parbox{3.5cm}{\raggedleft Mean zero-crossing period (spectral)}} \qquad & \overline{T}_{s,0} = \sqrt{\frac{m_0}{m_2}} \\
    \text{\parbox{3.5cm}{\raggedleft Mean zero-crossing period (direct)}} \qquad & \overline{T}_{d,0} = \frac{1}{N} \sum_{i=0}^N t_i
\end{align}

where $t_i$ refers to the zero-crossing periods of all waves in the corresponding surface elevation slice (zero-crossings determined by linear interpolation), and the expression for $\overline{T}_p$ is taken from \citet{young_determination_1995}.

For the characteristic wave steepness $\epsilon$ we use the peak wave number $k_p$, approximated from the peak period \eqref{eq:peak-period} and dispersion relation \eqref{eq:dispersion}, following \citet{serio_computation_2005}:

\begin{equation}
    \epsilon = \sqrt{2 m_0} k_p
\end{equation}

\subsubsection{Spectral Bandwidth and Benjamin-Feir Index}

The computation of spectral bandwidth follows \citet{serio_computation_2005}. As is the case with wave period, there is more than one way to estimate spectral bandwidth from data; in fact, there are at least 3 common quantities:

\begin{align}
\text{Broadness} \qquad & \sigma_B = \sqrt{1 - \frac{m_2^2}{m_0 \cdot m_4}} \nonumber \\
\text{Narrowness} \qquad & \sigma_N = \sqrt{\frac{m_0 \cdot m_2}{m_1^2} - 1} \label{eq:bandwidth} \\
\text{Peakedness} \qquad & \sigma_Q = \frac{m_0^2}{2 \sqrt{\pi}} \left(\int_0^\infty f \cdot \mathcal{S}(f)^2 \;\mathrm{d}f\right)^{-1} \nonumber
\end{align}

Some authors also refer to peakedness as ``quality factor''.

Broadness is problematic because of the occurrence of $m_4$, the fourth moment of the spectral density $\mathcal{S}$. Due to the $f^4$ term occurring in its estimation, broadness is extremely sensitive to the high-frequency tail of $\mathcal{S}$, which renders it an unacceptably noisy quantity at lower sampling rates (such as CDIP's \SI{1.28}{\hertz}). Therefore, FOWD only includes narrowness and peakedness as spectral bandwidth estimates.

The Benjamin-Feir index (BFI) was introduced in \citet{janssen_nonlinear_2003} and is a central parameter quantifying the strength of nonlinear interactions. Following \citet{serio_computation_2005}, we compute the BFI from steepness $\epsilon$, bandwidth $\sigma$ (which could be either of the 3 definitions above), peak wave number $k_p$, and depth $D$ as

\begin{equation}
    \operatorname{BFI} = \frac{\epsilon \nu}{\sigma} \sqrt{\max\{\beta / \alpha, 0\}}
\end{equation}

with

\begin{align}
    \nu =\;& 1 + \frac{2 k_p D}{\sinh(2 k_p D)} \\
    \alpha =\;& 2 - \nu^2 + 8 (k_p D)^2 \cdot \frac{\cosh(2 k_p D)}{\sinh^2(2 k_p D)}\\
    \beta =\;& \frac{8 + \cosh(4 k_p D) - 2 \tanh^2(k_p D)}{8 \sinh^4(k_p D)}  \nonumber \\
           &- \frac{\left(2 \cosh^2(k_p D) + \frac{\nu}{2} \right)^2}{\sinh^2(2k_p D) \cdot \left( \frac{k_p D}{\tanh(k_p D)} - \frac{\nu}{2} \right)^2}
\end{align}

In FOWD, we compute the BFI twice, with spectral bandwidth $\sigma$ estimated through both narrowness and peakedness (as defined in \eqref{eq:bandwidth}).

\subsubsection{Crest-trough Correlation}

\begin{figure}
    \centering
    \includegraphics[width=18pc]{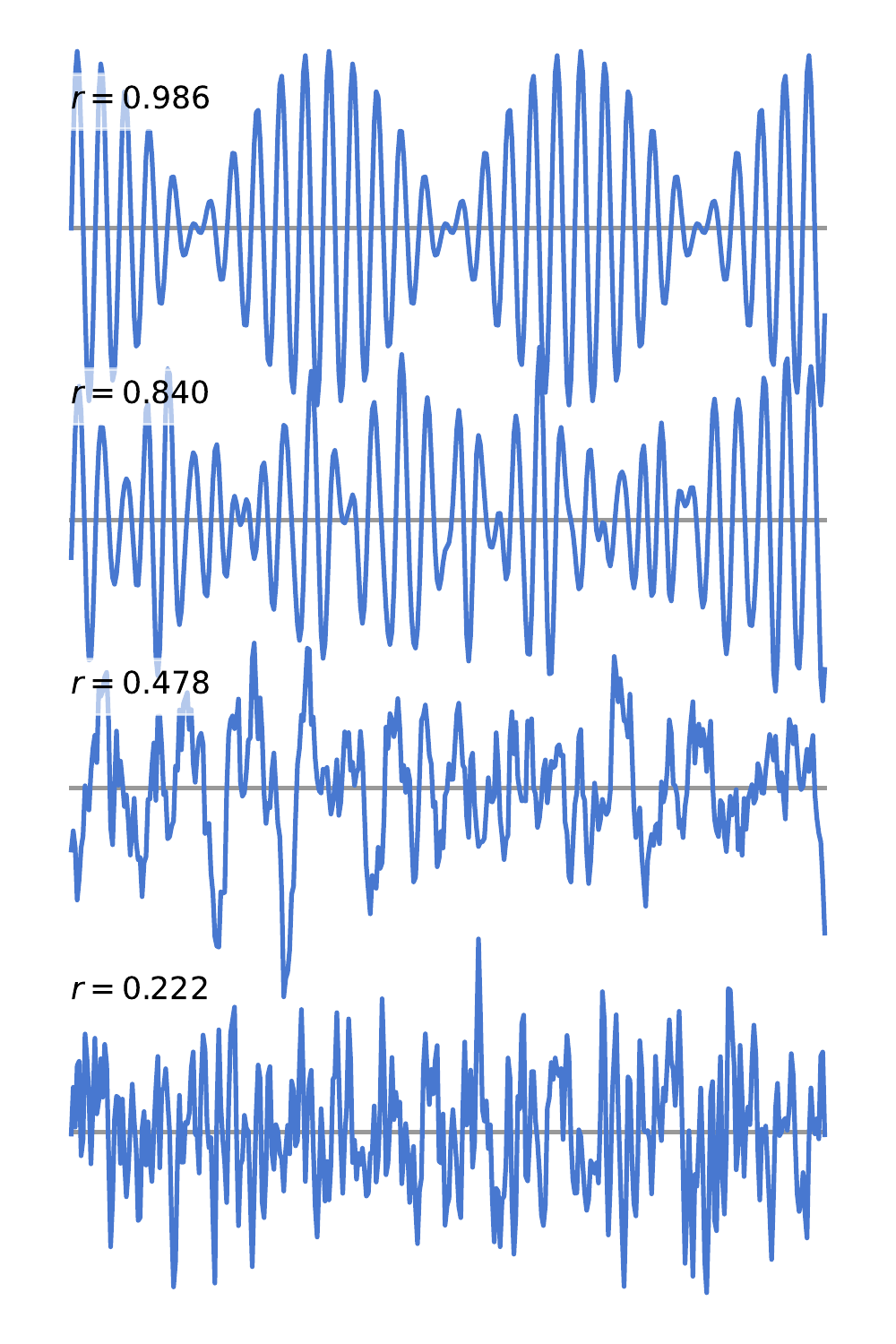}
    \caption{The crest-trough correlation $r$ is higher in ``groupy'', low-bandwidth sea states. Shown are surface elevations generated from Ochi-Hubble spectra \citep{ochi1977six} with increasing spectral bandwidth (from top to bottom), and the corresponding value of $r$.} \label{fig:crest-trough-corr}
\end{figure}

\citet{tayfun_m._aziz_distribution_1990} suggests another key parameter to describe wave height distributions, the correlation coefficient $r$ between squared crest height $A_0^2$ and squared trough depth $A_1^2$, which we refer to as ``crest-trough correlation''. $r$ is closely related to spectral bandwidth (as, for narrowband seas, crests and troughs are approximately of the same size, becoming increasingly chaotic / uncorrelated as more harmonics are added). By extension, it is also a measure for the tendency of the sea state to form wave groups (\figref{fig:crest-trough-corr}).

The estimation of crest-trough correlation from the spectral density $\mathcal{S}$ is further elaborated in \citet{tayfun_wave-height_2007}. Following these lines, we compute $r$ via:

\begin{equation}
    r = \frac{1}{m_0} \sqrt{\rho^2 + \lambda^2}
\end{equation}

with

\begin{align}
    \rho &= \int_0^\infty \mathcal{S}(\omega) \cos\left(\omega \frac{\overline{T}}{2}\right) \;\mathrm{d}\omega \\
    \lambda &= \int_0^\infty \mathcal{S}(\omega) \sin\left(\omega \frac{\overline{T}}{2}\right) \;\mathrm{d}\omega
\end{align}

where $\overline{T} = m_0 / m_1$ is the spectral mean period, and $\omega = 2\pi f$ the angular frequency.

\subsubsection{Spectral Partitioning}

To characterize processes that act mostly on short or long waves, spectral energy content is often more indicative than quantities based on the whole spectrum (such as mean period). Therefore, FOWD includes the relative energy content $\mathcal{E}$ over several spectral bands, computed as a definite integral over the spectral density $\mathcal{S}$:

\begin{equation}
    \mathcal{E}_i = \frac{\int_{f_{i}} \mathcal{S}(f) \;\mathrm{d}f}{\int_{0}^{\infty} \mathcal{S}(f) \;\mathrm{d}f} = \frac{1}{m_0} \int_{f_{i}} \mathcal{S}(f) \;\mathrm{d}f \label{eq:rel-energy}
\end{equation}

We use 5 distinct spectral bands (with limits $f_i$), each characteristic for a different physical regime (\tabref{tab:frequency-bands}).

(This is a crude way to perform spectral partitioning compared to more sophisticated approaches that take directionality into account \citep{portillayandun_climate_2016,portillayandun_global_2018}. However, this simple integral is straightforward to compute and interpret, and can be estimated using only a surface displacement time series.)

Similarly to the relative energy content, we also compute the total energy density contained in each frequency band (measured in \si{\joule\per\metre\squared}):

\begin{equation}
    P_i = \rho g \int_{f_{i}} \mathcal{S}(f) \;\mathrm{d}f
\end{equation}

with approximate density of sea water $\rho = \SI{1024}{\kilo\gram\per\metre\cubed}$, and gravitational acceleration $g = \SI{9.81}{\metre\per\second\squared}$.

\begin{table}
    \caption{Frequency bands used by FOWD and their approximate corresponding physical regime \citep[as e.g.\ given in][]{holthuijsen_waves_2010}.}
    \label{tab:frequency-bands}
    \centering
    \begin{tabular}{llp{3cm}}
        Band ID & Frequency range & Corresponding wave regime \\ \midrule
        1 & $< \SI{0.05}{\hertz}$ & Tides and seiches \\
        2 & \SIrange{0.05}{0.1}{\hertz} & Swell \\
        3 & \SIrange{0.1}{0.25}{\hertz} & Long-wave wind sea \\
        4 & \SIrange{0.25}{1.5}{\hertz} & Short-wave wind sea \\
        5 & \SIrange{0.08}{0.5}{\hertz} & Entire local wind-sea \\
        \bottomrule
    \end{tabular}
\end{table}

\subsubsection{Angular Integrals}

To make it possible to investigate the dependence of waves on phenomena like swell-wind sea crossing angles, we also split directional quantities into 5 distinct frequency bands, analogously to spectral energy content (\tabref{tab:frequency-bands}).

Since directional spread and wave direction are measured as an angle, we need to take special care when averaging these quantities. Furthermore, we want to weight the directional value at each frequency with the corresponding spectral energy at that frequency, to ensure that the resulting average represents the dominant angle within this frequency band.

To achieve this, we compute the integral of the directional quantity $q$ (which can be either dominant direction or directional spread) component-wise in Cartesian coordinates, weighted with the spectral density $\mathcal{S}$:

\begin{align}
\overline{x} &= \int_{f_i} \mathcal{S}(f) \sin q(f) \;\mathrm{d}f \\
\overline{y} &= \int_{f_i} \mathcal{S}(f) \cos q(f) \;\mathrm{d}f
\end{align}

where $f_i$ again demarcates the boundaries of each frequency band. Then, we transform the resulting Cartesian components back to an angle:

\begin{equation}
\overline{q} = \arctan\left(\frac{\overline{x}}{\overline{y}} \right)
\end{equation}

which is the desired weighted angular average.

\subsubsection{Directionality Index}

A key parameter to characterize the influence of directional spread on the wave dynamics is the ``directionality index'' $R$ \citep[as introduced in][]{fedele_kurtosis_2015}. It is commonly defined as:

\begin{equation}
    R = \frac{\sigma_\theta^2}{2 \nu^2}
\end{equation}

where $\sigma_\theta$ is the directional spread (in radians), and $\nu$ denotes the spectral bandwidth \citep[we use narrowness, as in][]{fedele_large_2019}.

This factor $R$ makes it possible to compute various directionality-corrected versions of e.g.\ the Benjamin-Feir index and kurtosis \citep{fedele_kurtosis_2015,fedele_large_2019}.

In FOWD, we estimate $R$ by computing the narrowness of the spectrum as provided by CDIP. Directional spread is computed as outlined above, which we integrate over all frequencies to obtain $\sigma_\theta$.

\subsection{Running Window Processing} \label{sec:window}

Usually, studies that investigate extreme wave observations divide all data into blocks of equal length in time, e.g.\ \SI{30}{\minute} chunks, that are then analyzed separately \citep[e.g.][]{casasprat_short-term_2010,cattrell_can_2018}. However, the transient nature of the ocean has long been identified as a potential source for systematic error \citep{adcock_physics_2014,gemmrich_dynamical_2011,gemmrich_spatial_2016}, as it is not clear that the wave height distribution is the same for the first and the last wave within each chunk.

A related consideration is that the estimated quantities must be \emph{agnostic of the future} --- i.e., look-aheads must be impossible. This property is critical for machine learning applications, where future state leaking into the training data may completely invalidate the generalization abilities of a machine learning algorithm.

We have therefore decided to use a running window approach in FOWD. Here, we iterate through the raw data one zero-upcrossing at a time, computing the characteristic sea state parameters based on the immediate history of every wave. This implies that there is no time gap between the end of the aggregation period and the current wave, at the expense of additional computation time (since the sea state has to be re-computed for every wave).

\begin{figure}
    \centering
    \includegraphics[width=18pc]{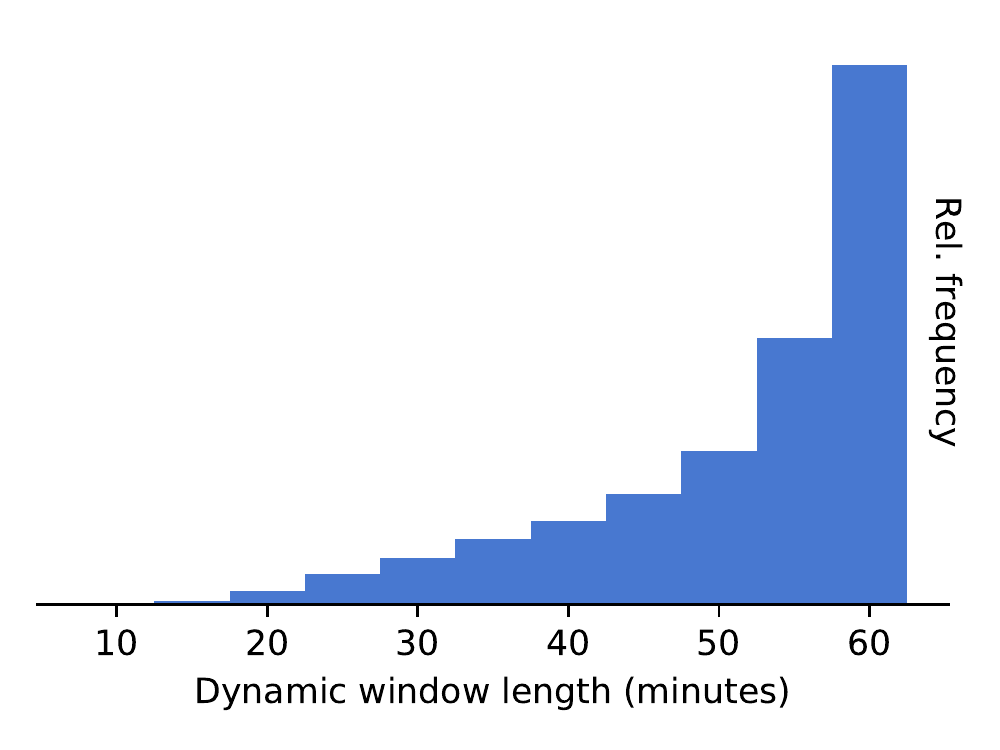}
    \caption{Most dynamic windows are longer than \SI{30}{\minute}. Shown is a histogram of the determined optimal window size across all Hawaiian CDIP stations.} \label{fig:dynamic-window}
\end{figure}

Picking a window length is always a trade-off between bias (longer windows are more prone to non-stationarity) and variance (shorter windows leave us with less data to work with). Therefore, all parameters are computed 3 times:

\begin{itemize}
    \item Twice using fixed \SI{30}{\minute} and \SI{10}{\minute} windows. This makes it possible to investigate the stationarity of the current sea state by comparing the values obtained from each window length.

    \item One more time using a variable, data-dependent window as suggested in \citet{boccotti_wave_2000} and used in \citet{fedele_large_2019}. We define the optimal window size $n$ to be the one that minimizes

    \begin{equation}
        \operatorname{Std}\bigg(\frac{\sigma_{n,i+1}}{\sigma_{n,i}} - 1\bigg) \label{eq:dynamic-window}
    \end{equation}

    where $\sigma_{n,i}$ is the standard deviation of the sea surface elevation in the $i$-th chunk with length $n$, applied to the past \SI{12}{\hour} of time series.

    To make this process more robust, we re-compute \eqref{eq:dynamic-window} 10 times for each candidate window with a different time offset. FOWD tries a total of 11 different windows lengths between \SI{10}{\minute} and \SI{60}{\minute} and selects the one that minimizes the sum of \eqref{eq:dynamic-window} across all trials.

    This process tends to generate time windows longer than \SI{40}{\minute} in most conditions, but is also capable of reducing the window size if needed (\figref{fig:dynamic-window}).

    As the standard deviation of the sea surface elevation $\sigma$ is directly related to significant wave height, we expect this to yield near-optimal window sizes for significant wave height and other slowly drifting quantities (such as mean period and energy content), but suboptimal results for faster drifting parameters (such as steepness, peak period, kurtosis).
\end{itemize}

\subsection{Quality Control} \label{sec:qc}

FOWD uses a combination of QC flags, most of which are inspired by the process suggested in \citet{christou_field_2014}. A measurement is discarded if any of the following conditions are met when applied to the past \SI{30}{\minute} surface elevation:

\begin{enumerate}
    \item[(a)] Any waves with zero-crossing period $> 25$s.
    \item[(b)] The rate of change of the surface elevation $\eta$ exceeds the limit rate of change by a factor of 2 or more at any point, i.e.,

    \begin{equation}
        \left|\frac{\partial\eta}{\partial t}\right| > 2 U_\text{lim}.
    \end{equation}

    The limit rate of change $U_\text{lim}$ is defined as

    \begin{equation}
        U_\text{lim} = 2 \pi \frac{\operatorname{Std}(\eta)}{\langle T_{d,0} \rangle} \cdot \sqrt{2 \ln N}
    \end{equation}

    with standard deviation $\operatorname{Std}$, mean observed zero-crossing periods $\langle T_{d,0} \rangle$, and number of waves in the record $N$.

    This criterion removes records containing waves that are much steeper than the average rate of change $\operatorname{Std}(\eta) /\langle T_{d,0} \rangle$, i.e.\ records with single, very steep waves, but leaves sea states with many steep waves intact.
    \item[(c)] 10 consecutive data points of the same value.
    \item[(d)] Any absolute crest or trough elevation that is greater than 8 times the normalized median absolute deviation (MADN) of the surface elevation, i.e.,

    \begin{equation}
        |h| > 8 \kappa \operatorname{median}(|\eta - \operatorname{median}(\eta)|)
    \end{equation}

    with $\kappa = 1.483$, which ensures that MADN converges to standard deviation for Gaussian distributed $\eta$ with growing sample size \citep[see e.g.][]{Huber2011}.

    This criterion permits crest heights and trough depths of up to about 2 times the significant wave height, which should be more than enough for any real signal. (In a linear sea, a crest exceeding $2 H_S$ would have a probability of $\exp(-32) \approx 10^{-14}$.)
    \item[(e)] Surface elevations are not equally spaced in time (but they may contain \texttt{NaN} values).
    \item[(f)] The ratio of missing (\texttt{NaN}) to valid data exceeds \SI{5}{\percent}.
    \item[(g)] Less than 100 individual zero-crossings.
\end{enumerate}

All waves that fail QC and are larger than 2 times the significant wave height are written to a log file to allow for manual inspection. Additionally, all waves that are larger than 2.5 times the significant wave height are written to the log file, regardless whether they pass QC or not. This enables us to evaluate the QC process and tweak thresholds or exclude faulty sub-datasets as needed.

A brief evaluation of this QC process when applied to the CDIP data is given in \subsecref{sec:cdip}{sec:qc-cdip}.

\subsection{Additional Metadata \& Reproducibility} \label{sec:reproducibility}

All FOWD output files are self-documenting in the sense that they include all relevant metadata as netCDF4 attributes, both for each variable and the dataset as a whole.

Apart from the static metadata documenting the coordinates and parameters (that is the same for every FOWD output file), we also include some metadata related to the processing environment and raw data source to ensure reproducibility.

Specifically, each wave record includes the time stamp, file name and UUID of the raw source file it came from (see \quantitytable). The output files also include the exact version of the FOWD processing implementation used to create the file in form of a git tag, along with a unique file identifier (UUID).

That way, we enable users to reproduce any result by allowing them to use the exact same processing version and input file.

\section{Reference Implementation} \label{sec:implementation}

As part of this work, we supply a Python reference implementation of the FOWD processing toolkit. It makes use of the popular Python packages \verb|xarray|, \verb|numpy| and \verb|scipy| to process large amounts of input data efficiently. The implementation processes either CDIP netCDF4 files or generic input files in a fixed netCDF4 format. Multiple CDIP deployments (within the same station) can be processed in parallel.

\subsection{Memory Efficiency}

Due to FOWDs running-window approach (see \subsecref{sec:specification}{sec:window}), FOWD output datasets are about 10 times bigger than the input surface elevation time series (since every wave results in about 80 output features). This demands that the processing implementation does not store entire output files in memory, but instead continually writes results to disk.

We achieve this by keeping only the immediate \SI{30}{\minute} history of the current processing time in memory. Each new record is flushed to disk using Python's pickle format. After the processing has finished, these pickle files are read back by the main process in chunks, reformatted to the netCDF4 output format, and flushed to disk again. This ensures that the main process uses only a negligible amount of memory, while each worker process only keeps the input data in memory. In other words, if the input data fits in memory, processing will succeed.

\subsection{Testing Strategy}

\begin{figure*}
    \centering
    \includegraphics[height=.92\textheight]{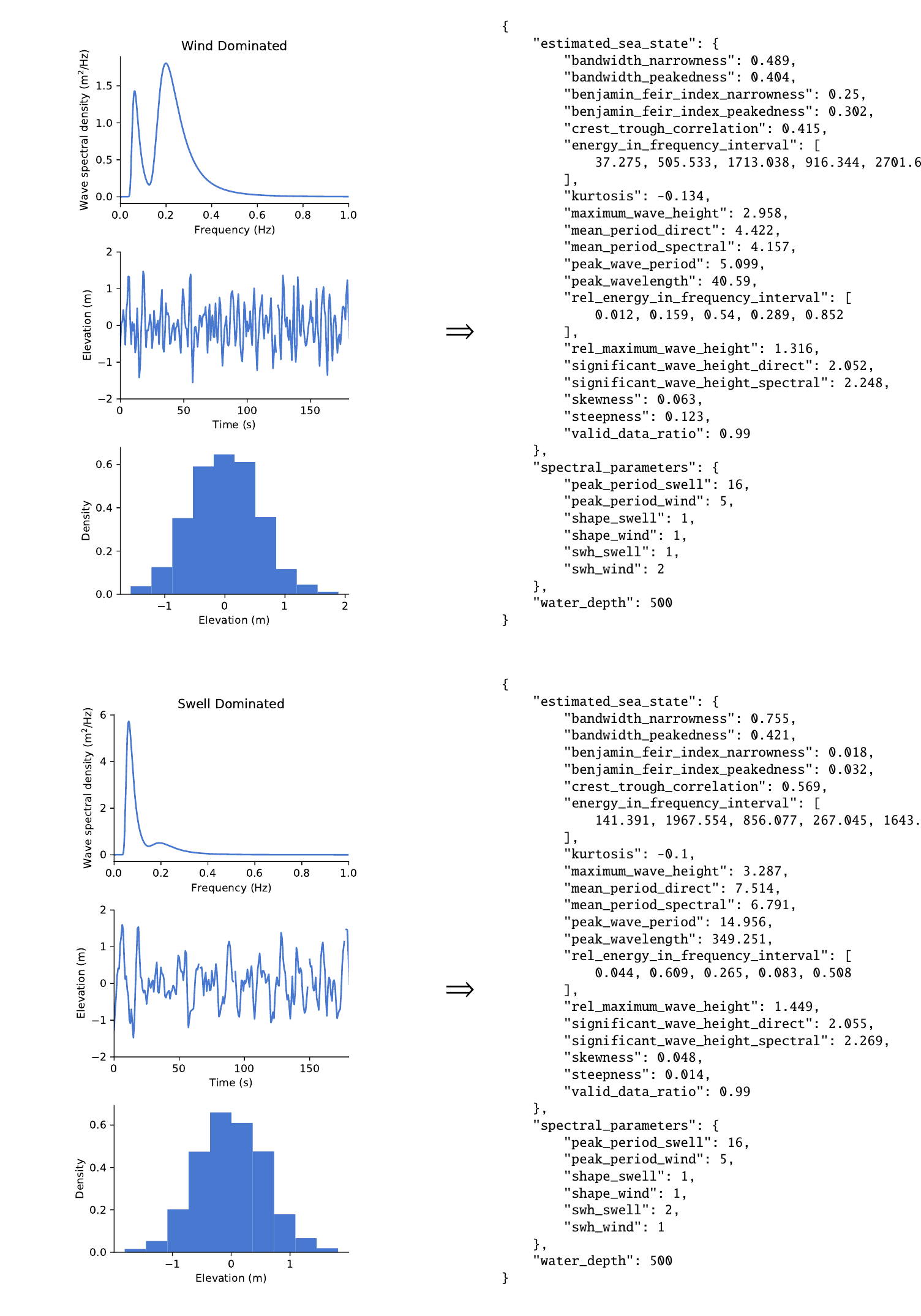}
    \caption{Sanity check test cases allow us to verify manually that computed parameters are reasonable. Shown are test inputs (left) and outputs (right) for high-frequency (top) and low-frequency (bottom) seas. Estimated sea state parameters are defined in \quantitytable. Spectral parameters are input parameters of the Ochi-Hubble spectrum used to generate each test case (as shown in upper left panels).} \label{fig:sanity}
\end{figure*}

In software engineering, automated tests are an invaluable tool to ensure proper functionality of a product. Unfortunately, writing automated tests for processing workflows of physical data is often impossible or infeasible due to the lack of ground-truth answers to compare to. On the other hand, faulty results are often easy to detect for humans when they fall outside of reasonable physical limits or show the wrong scaling behavior. We have therefore opted for semi-automated \emph{sanity checks} instead of fully automated unit tests for the core processing.

Each sanity check test case generates a random surface elevation time series from a different ground-truth wave spectrum and runs it through the FOWD processing. Here, only the spectral shape is prescribed externally, surface elevations are drawn as harmonics with random phases from the spectrum. The resulting output parameters can then be inspected manually.

Two example sanity check spectra are bi-modal Ochi-Hubble spectra \citep{ochi1977six} that are either swell dominated (low frequency peak is dominant) or wind dominated (high frequency peak is dominant). We would expect that the wind dominated spectrum leads to lower period, higher steepness and BFI, and shorter wavelength. In both cases, we expect to find a spectral significant wave height of

\begin{equation}
    \operatorname{SWH}_\text{total} = \sqrt{\operatorname{SWH}_\text{swell}^2 + \operatorname{SWH}_\text{wind}^2}
\end{equation}

and excess kurtosis and skewness around $0$. Directly estimated significant wave height ($H_{1/3}$) is usually slightly lower than its spectral counterpart ($H_{m_0}$), and vice-versa for wave period.

Indeed, all of these expectations are met for this particular test case (Fig.~\ref{fig:sanity}). Other sanity checks feature idealized spectra, e.g\ containing just a single harmonic, that allow us to validate parameters that are more difficult to interpret like crest-trough correlation, or idealized directional spectra (not shown).

Thanks to these sanity checks, we are confident that the FOWD core processing is free from critical errors.

\section{Processing of CDIP Buoy Data} \label{sec:cdip}

The following sections describe the CDIP input and FOWD output data, analyze QC performance and the impact of FOWD's running window processing, and discuss some caveats that apply when using buoy data for extreme wave studies.

\subsection{Input Data and Processing}

In total, the CDIP catalogue spans about 750 years of continuous surface elevation measurements (almost all at sampling rates of \SI{1.28}{\hertz}) that are available in netCDF4 format through a THREDDS server. This amounts to about \SI{270}{\giga\byte} of raw data.

While CDIP data files also include horizontal displacements and a number of derived quantities (like significant wave height, peak period, \ldots), we use only the raw vertical surface displacement, station metadata, and directional quantities for processing. This ensures that FOWD is applicable to any instrument that delivers a surface displacement time series (including radar or laser sensors).

We applied only minimal preprocessing to the data, which consists of removing all data that has an error flag set and subtracting the \SI{30}{\minute} running mean from the raw vertical surface elevation. After that, we processed all data (using the FOWD reference implementation described in \secref{sec:implementation}) in about 72 hours on 10 cluster nodes in parallel. The resulting output dataset has a total (compressed) size of \SI{1.1}{\tera\byte}. We create one output file per CDIP station, with individual file sizes ranging between \SI{1.7}{\mega\byte} and \SI{38}{\giga\byte}. In total, FOWD contains about $4.2$ billion individual waves and sea states.

\subsection{Quality Control and Filtering} \label{sec:qc-cdip}

As outlined in \subsecref{sec:specification}{sec:qc}, FOWD automatically logs waves failing QC that are higher than 2 significant wave heights, and all waves higher than 2.5 significant wave heights (whether they pass QC or not) in JSON format. This allows us to assemble some higher-order statistics to get an idea of how prevalent quality issues are in the CDIP data, and to verify that FOWD's QC system works as intended.

\begin{table}
    \centering
    \caption{The number of times each QC flag was triggered for the whole CDIP catalogue. See \subsecref{sec:specification}{sec:qc} for a definition of flags a--g. Note that multiple flags can be active for the same wave.} \label{tab:qc-flags}
    \begin{tabular}{p{1.2cm}r}
        Flag & Times fired \\ \midrule
        a & \num{31547} \\
        b & \num{18465} \\
        c & \num{39470} \\
        d & \num{47544} \\
        e & \num{0} \\
        f & \num{11915} \\
        g & \num{4089} \\ \midrule
        Failed waves & \num{77371}
    \end{tabular}
\end{table}

In total, just under \num{80000} waves fail QC (\tabref{tab:qc-flags}). About \SI{80}{\percent} of these QC failures occur in only 5 CDIP locations (out of 161). This suggests that relatively few deployments with general quality problems cause a majority of QC failures.

To investigate this further and isolate faulty deployments, the FOWD implementation includes a post-processing command that produces plots of all records in the QC logs. These plots show the raw surface elevation of the failing wave and its immediate \SI{30}{\minute} history.

After inspecting each of these plots, we decided to blacklist 38 deployments and 4 entire CDIP stations that showed obvious quality problems like frequent spikes, extreme oscillations, unphysical values, or jumps (\tabref{tab:blacklist}).

\begin{table}
    \centering
    \caption{Blacklisted CDIP deployments that failed visual inspection.} \label{tab:blacklist}
    \small
    \begin{tabular}{lp{3cm}}
        CDIP ID & Excluded deployments \\ \midrule
        045p1 & d01, d02, d03, d13, d15, d17, d19, d21 \\
        094p1 & d01, d02, d03, d04, d05 \\
        096p1 & d04 \\
        100p1 & d11 \\
        106p1 & d02 \\
        109p1 & d05, d06 \\
        111p1 & d06 \\
        132p1 & d01 \\
        141p1 & d03 \\
        142p1 & d02, d15, d18 \\
        144p1 & d01 \\
        146p1 & d01, d02 \\
        158p1 & d02, d04 \\
        162p1 & d07 \\
        163p1 & d01, d05 \\
        167p1 & d01 \\
        172p1 & d01 \\
        177p1 & all deployments \\
        196p1 & d04 \\
        201p1 & d03 \\
        205p1 & all deployments \\
        206p1 & all deployments \\
        261p1 & all deployments \\
        430p1 & d06 \\
        431p1 & d02 \\ \bottomrule
    \end{tabular}
\end{table}

On top of excluding these blacklisted CDIP deployments, we also removed all records in conditions where buoys are known to be unreliable \citep[inspired by][]{mcallister_experimental_2019}:

\begin{enumerate}
    \item Records with \SI{30}{\minute} significant wave height smaller than \SI{1}{\metre}.
    \item Records with spectral mean frequency higher than $1/3.2$ of the Nyquist frequency. For \SI{1.28}{\hertz} data, this is equivalent to filtering all records with a mean wave period below \SI{5}{\second}.
    \item Records where the relative energy content of frequency band 1 exceeds \SI{10}{\percent} (extensive low-frequency drift).
\end{enumerate}

After filtering, the final dataset contains about 1.4 billion waves and sea states (about \SI{67}{\percent} filtered, most due to the minimum significant wave height requirement).

Since FOWD is also intended for use by non-wave experts, it is essential to provide access to a pre-cleaned dataset. Therefore, the filtered FOWD-CDIP dataset is available for download along with the unfiltered one (see data availability statement).

\subsection{Impact of Running Window Processing}

\begin{figure*}
    \centering
    \includegraphics[width=.8\textwidth]{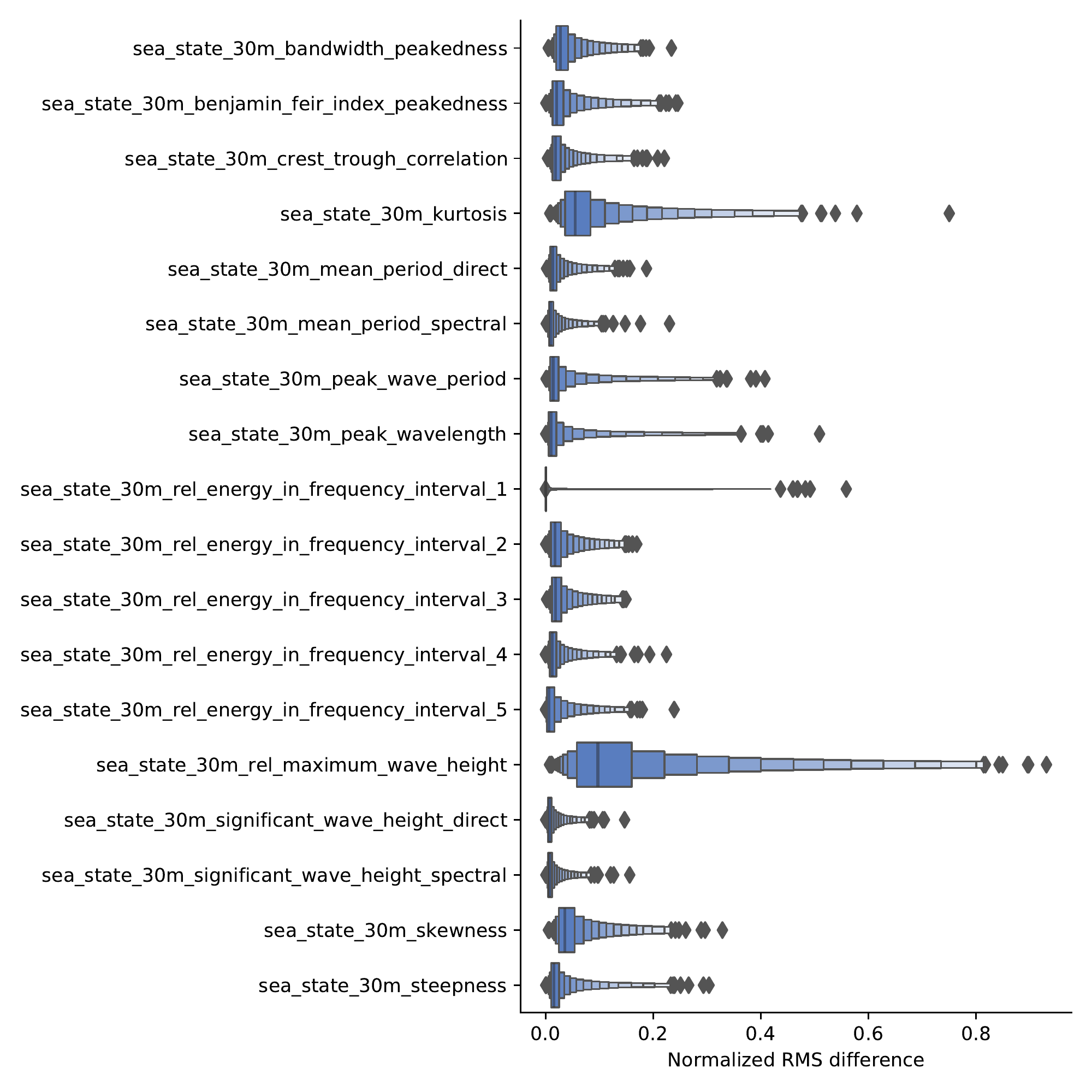}
    \caption{In extreme cases, using running windows (instead of fixed chunks) leads to RMS differences of up to \SI{50}{\percent} of the characteristic scale of a parameter (\tabref{tab:param-scale}). Shown is the distribution of normalized RMS difference between processing based on running windows and fixed chunks for some parameters.}
    \label{fig:window-rms}
\end{figure*}

After processing the CDIP data, we can now investigate how large of a difference FOWD's running window processing (as described in \subsecref{sec:specification}{sec:window}) makes in practice, compared to the usual fixed-window approach.

To this end, we divide the FOWD catalogue for one particular CDIP station (with ID 188p1, containing about 30 million waves) into \SI{30}{\minute} chunks. The last measurement in each of these chunks (concerning the past \SI{30}{\minute} sea state) then represents what would have been obtained for all waves if FOWD did not use running windows.

We can then quantify the influence of the running window approach by computing the root mean square (RMS) difference between this last measurement of every chunk and all other data points in it. To make it easier to compare the different parameters, we divide each by a characteristic scale to obtain a normalized RMS (\tabref{tab:param-scale}).

\begin{table*}
    \caption{Characteristic scale used to normalize root-mean-square residual for each parameter (\figref{fig:window-rms}).}
    \label{tab:param-scale}
    \centering
    \footnotesize
    \begin{tabular}{lll}
        Parameter & Typical range & Resulting scale \\ \midrule
        \param{sea\_state\_30m\_bandwidth\_peakedness} & 0 -- 0.6 & 0.6 \\
        \param{sea\_state\_30m\_benjamin\_feir\_index\_peakedness} & 0 -- 0.6 & 0.6 \\
        \param{sea\_state\_30m\_crest\_trough\_correlation} & 0.2 -- 1.0 & 0.8 \\
        \param{sea\_state\_30m\_kurtosis} & -0.5 -- 1.5 & 2.0 \\
        \param{sea\_state\_30m\_mean\_period\_direct} & \SI{4}{\second} -- \SI{15}{\second} & \SI{11}{\second} \\
        \param{sea\_state\_30m\_mean\_period\_spectral} & \SI{4}{\second} -- \SI{15}{\second} & \SI{11}{\second} \\
        \param{sea\_state\_30m\_peak\_wave\_period} & \SI{4}{\second} -- \SI{20}{\second} & \SI{16}{\second} \\
        \param{sea\_state\_30m\_peak\_wavelength} & \SI{0}{\metre} -- \SI{600}{\metre} & \SI{600}{\metre} \\
        \param{sea\_state\_30m\_rel\_energy\_in\_frequency\_interval\_1} & 0 -- 0.2 & 0.2 \\
        \param{sea\_state\_30m\_rel\_energy\_in\_frequency\_interval\_2} & 0 -- 1 & 1 \\
        \param{sea\_state\_30m\_rel\_energy\_in\_frequency\_interval\_3} & 0 -- 1 & 1 \\
        \param{sea\_state\_30m\_rel\_energy\_in\_frequency\_interval\_4} & 0 -- 0.4 & 0.4 \\
        \param{sea\_state\_30m\_rel\_energy\_in\_frequency\_interval\_5} & 0 -- 1 & 1 \\
        \param{sea\_state\_30m\_rel\_maximum\_wave\_height} & 1.2 -- 2.2 & 1 \\
        \param{sea\_state\_30m\_significant\_wave\_height\_direct} & \SI{0.5}{\metre} -- \SI{8.0}{\metre} & \SI{7.5}{\metre} \\
        \param{sea\_state\_30m\_significant\_wave\_height\_spectral} & \SI{0.5}{\metre} -- \SI{8.0}{\metre} & \SI{7.5}{\metre} \\
        \param{sea\_state\_30m\_skewness} & -0.5 -- 0.5 & 1 \\
        \param{sea\_state\_30m\_steepness} & 0 -- 0.12 & 0.12 \\
        \bottomrule
    \end{tabular}
\end{table*}

The resulting distribution of the normalized RMS in each chunk shows that, while deviations are typically below \SI{10}{\percent} of the characteristic scale, they can reach up to \SI{50}{\percent} in extreme cases (\figref{fig:window-rms}). As expected, some parameters (such as kurtosis, maximum wave height) are much more prone to drift than others (such as significant wave height, spectral energy). However, this is sensitive to which characteristic scale we choose, so comparisons between parameters remain noisy.

A particularly important quantity in this context is the significant wave height. If the significant wave height is underestimated with an error of only 5\%, a wave with true abnormality index $\abni=2$ is estimated as a wave with $\abni = 2.1$, which is less than half as likely to occur (assuming Rayleigh distributed waves).

We conclude that the running window approach \emph{can} lead to significantly different results, apart from the more important effect of preventing look-aheads (as discussed in \subsecref{sec:specification}{sec:window}). In other words, explicitly accounting for a drifting sea state provides an opportunity to reduce bias by a non-trivial amount --- although we did not measure how much this approach influences final results or conclusions.

\subsection{Shortcomings of Buoy Data} \label{sec:shortcomings}

Although any dataset that provides surface elevation measurements can be processed into a FOWD dataset, buoy data remains a dominant data source due to its relatively large availability (at least compared to radar and laser measurements). Therefore, this section discusses some of the known problems with buoy data, and how they carry over to FOWD and its possible applications.

First and foremost, buoys tend to linearize surface elevations to some degree \citep[see][for a discussion]{mcallister_experimental_2019,mcallister_lagrangian_2019}. This is especially problematic in rough seas with high steepness, because buoys can penetrate through a steep crest or move laterally around it and underestimate the true wave height. Combined with the inherent sampling variability of a point measurement \citep[the two-dimensional wave has to hit the buoy at the crest to be registered at full height, see][]{benetazzo_observation_2015}, wave estimates based on buoy data tend to be too conservative \citep[see also][]{casasprat_short-term_2010}.

This is of course inconvenient for studies with the goal to estimate absolute rogue wave risk, since one needs to take additional steps to correct for these biases, include other data sources, or accept that the results represent a lower bound for rogue wave risk.

However, this is not a problem when estimating the \emph{relative} importance of sea state risk factors, as buoys should be similarly inaccurate across a wide range of different sea states (after the most problematic conditions are filtered, see \subsecref{sec:cdip}{sec:qc-cdip} --- perhaps with the exception of very steep seas). We therefore see no problem with using buoy data for the type of study presented in \secref{sec:application}.

Another issue to keep in mind is \emph{selection bias}. Buoys tend to be placed in locations that are easy to reach and of special interest for humans. This results in coastal areas to be over-represented, so there might be less available data to study the conditions on the open ocean.

Finally, no reasonable amount of one-dimensional time series data can tell us about truly exceptional events. In offshore engineering contexts, an important quantity is the ``\num{10000} year wave'', which is the largest expected wave in a \num{10000} year period. Events of this rarity cannot be estimated with this dataset without additional work (such as further theoretical assumptions, or data augmentation via simulations).

\section{Example Application: Which Sea State Parameter is the Best Predictor for Rogue Wave Occurrence?} \label{sec:application}

As an example application of FOWD, we look at the connection between sea state and the occurrence of rogue waves to find which sea state parameter is the best predictor for rogue wave activity (where we find the largest change in rogue wave probability when varying the parameter).

In this context, we define rogue waves as any wave whose height exceeds twice the significant wave height, i.e. $\abni > 2$. For any given sea state with wave height distribution $P(\abni)$ we would expect the next wave to be a rogue wave with probability

\begin{equation}
    p = \int_{2}^{\infty} P(\abni) \; \mathrm{d}\abni
\end{equation}

From linear superposition of random waves with narrow spectral bandwidth \citep{longuet1952statisticaldistribution}, we would expect this criterion to be fulfilled for roughly 1 in \num{3000} waves.

In the filtered FOWD-CDIP dataset, this criterion is fulfilled for about \num{100000} out of 1.5 billion total waves (i.e., 1 in \num{15000}), with a significant amount of rogue waves occurring within seconds of one another (\tabref{tab:rogue-stats}).

This implies that the measured incidence rate of rogue waves across all sea states is about 5 times lower than predicted by linear theory. This is not uncommon for buoy data \citep{casasprat_short-term_2010}, and could to some degree be due to the underestimation of extreme waves by buoys (as discussed in \subsecref{sec:cdip}{sec:shortcomings}). However, we suspect that this is mostly caused by physical factors. Effects like crest-trough correlations $< 1$ (as we will see below) or wave breaking can severely limit the formation of rogue waves, and are not accounted for in linear theory.

\begin{table}
    \caption{Number of waves in the FOWD-CDIP dataset fulfilling various criteria.} \label{tab:rogue-stats}
    \centering
    \begin{tabular}{p{5cm}r}
        \toprule
        Waves with $AI < 2$ & \num{1383488167} \\
        Waves with $AI \geq 2$ & \num{82058} \\
        Waves with $AI \geq 2.2$ & \num{11849} \\
        Waves with $AI \geq 2.5$ & \num{564} \\
        Waves with $AI \geq 2$ within \SI{30}{\second} & \num{2455} \\
        \bottomrule
    \end{tabular}
\end{table}

During the following sections, we will take a closer look under which conditions rogue waves preferably occur. For this, we use the combined data from all Hawaiian CDIP stations (stations with IDs 098p1, 106p1, 146p1, 165p1, 187p1, 188p1, 198p1, 225p1, 233p1), containing about 200 million waves.

\subsection{Confounding and Roguish Sea States}

To get a feeling for the data, we investigate correlations between some of the sea state parameters, and have a look at the probability density functions of sea states in which we find rogues with $\abni > 2$ and $\abni > 2.4$.

\begin{figure*}
    \centering
    \includegraphics[width=.8\textwidth]{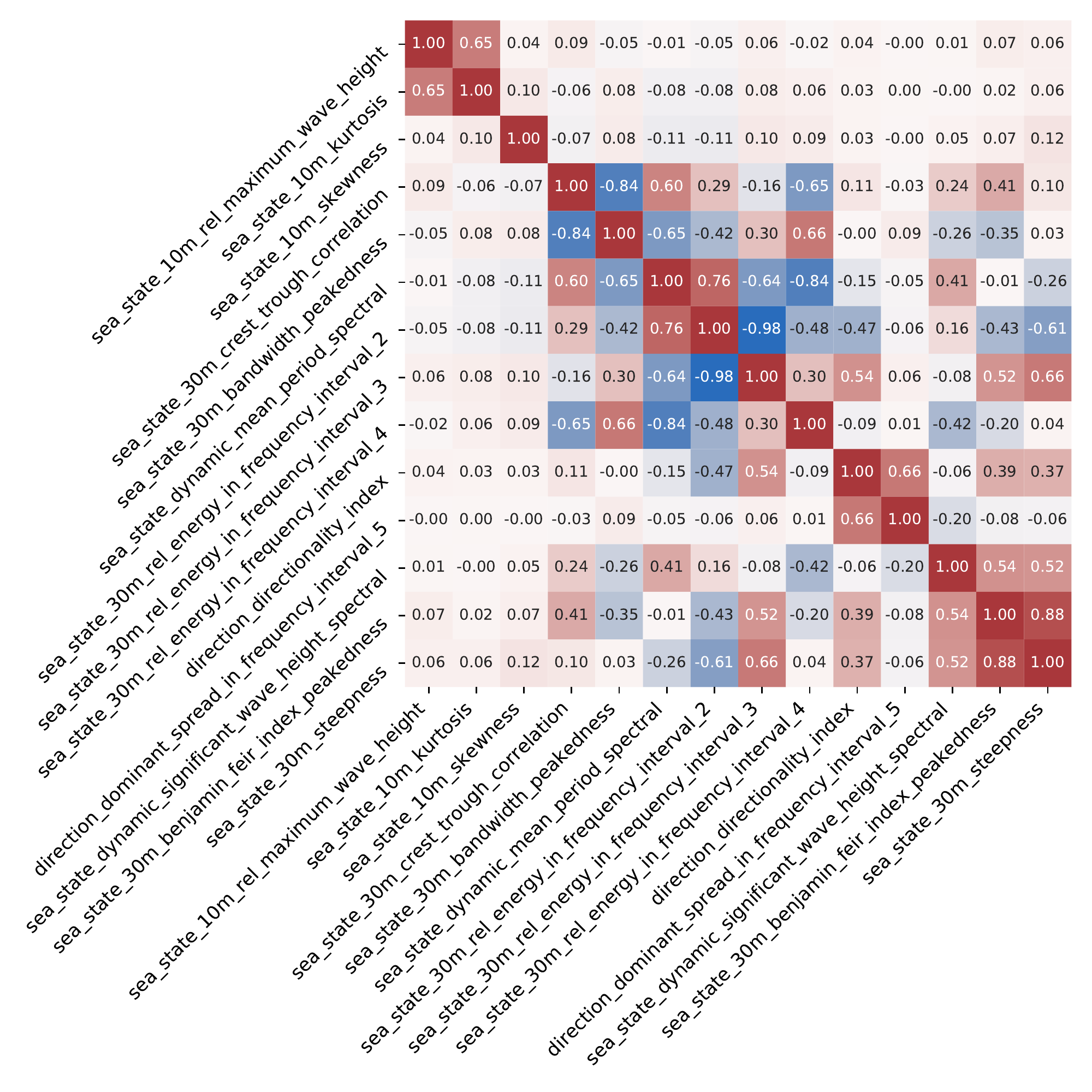}
    \caption{Linear (Pearson) correlation matrix of selected parameters. Almost all parameters are strongly correlated with at least one other parameter, but exceptions exist (skewness, kurtosis / maximum wave height, wind sea directional spread).}
    \label{fig:hawaii-correlation}
\end{figure*}

\begin{figure*}
    \centering
    \includegraphics[height=.9\textheight]{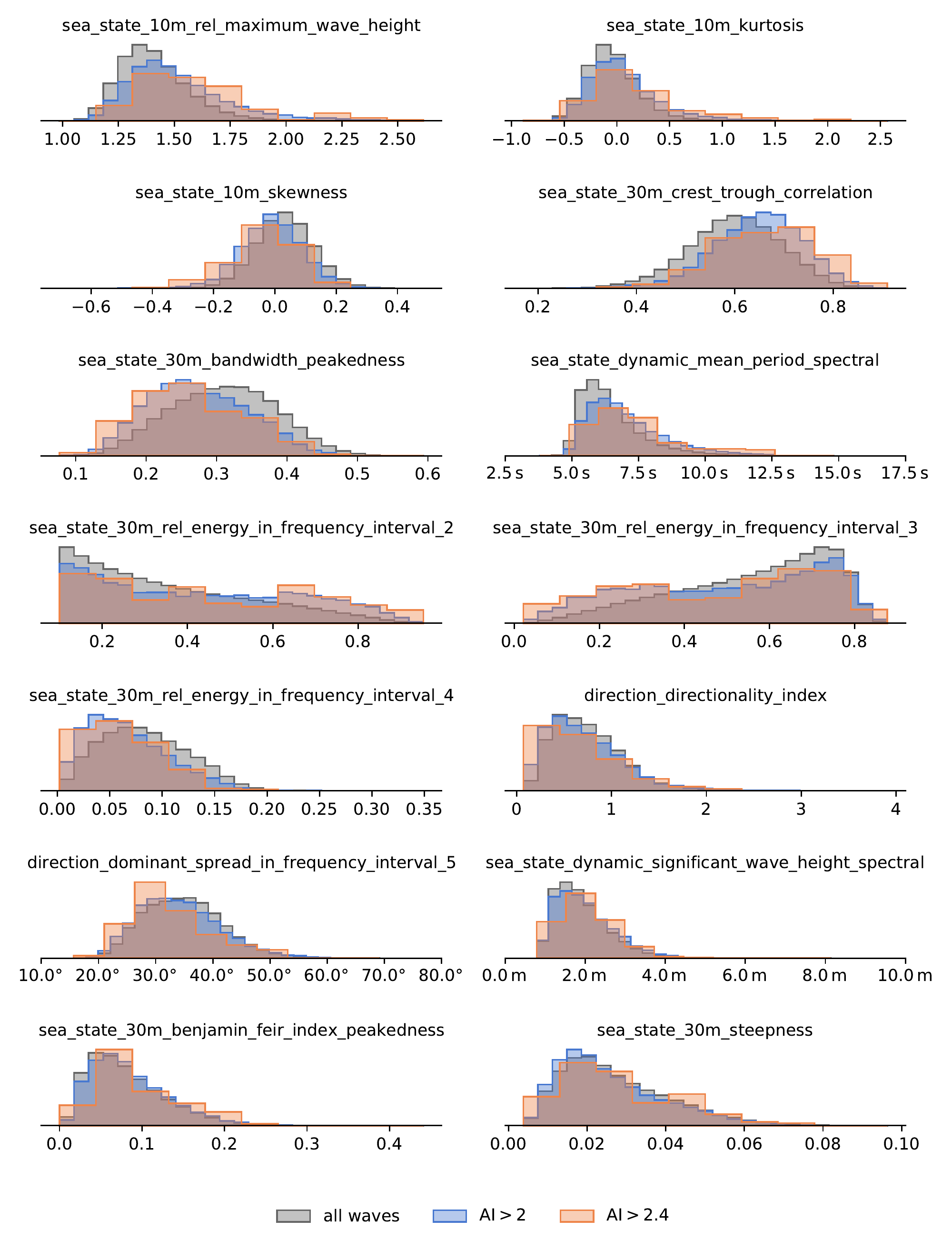}
    \caption{Most parameters show a clear difference between the probability distributions of all sea states and those containing an extreme wave, but some just show a weak dependence (e.g. directional spread, significant wave height, steepness). Shown are the probability density functions (PDFs) of various sea state parameters, estimated via histograms. Each parameter includes PDFs for the sea states of all waves, waves with $\abni > 2$, and waves with $\abni > 2.4$.}
    \label{fig:hawaii-feature-dist}
\end{figure*}

The correlation matrix of the sea state parameters (\figref{fig:hawaii-correlation}) provides yet another important sanity check for FOWD, since many parameters are correlated by definition (such as BFI, which is computed based on steepness and spectral bandwidth).

Furthermore, it serves as an important reminder that there are many non-obvious correlations, such as the one between spectral bandwidth and mean period. Any conclusion we draw about the influence of a parameter on rogue wave activity thus has to take possible confounders into account.

The probability density functions of roguish seas (\figref{fig:hawaii-feature-dist}) indicate several potential controlling parameters for rogue wave occurrence, where the distribution of seas containing a rogue wave differs substantially from that of all waves (with e.g.\ skewness, spectral bandwidth, and maximum wave height being promising candidates).

This analysis, while intuitively approachable, yields little quantitative insight into the relative importance of each parameter, and neglects the influence of sample size effects. The following section addresses this through a simple analytical Bayesian parameter estimation.

\subsection{Estimation of Rogue Wave Probabilities with Uncertainties}

A major challenge when dealing with rare events like rogue waves is to determine whether there actually is enough data to make a statement. We will therefore quantify this uncertainty through Bayesian credible intervals on the rogue wave probability $p$.

As the first step, we assume that the occurrence of $n^+$ rogue waves and $n^-$ non-rogue waves in a given sea state is drawn randomly with some rogue wave probability $p$. $n^+$ then follows a binomial distribution:

\begin{equation}
    n^+ \sim \operatorname{Binom}(n^+ + n^-,\; p)
\end{equation}

The goal of this analysis is to estimate $p$ from measurements of $n^+$ and $n^-$. For $p$, we encode prior information by assuming a Beta prior, given by

\begin{equation}
    p_\text{prior} \sim \operatorname{Beta}(\alpha_0,\; \beta_0)
\end{equation}

with parameters $\alpha_0$, $\beta_0$, which we choose as $\alpha_0 = 1$ and $\beta_0=10000$, roughly representing the expected order of magnitude $\mathcal{O}(p) \approx 10^{-4}$ (this is just a weakly informative prior to constrain $p$ to the right order of magnitude --- the exact values have no influence on the conclusions of this analysis).

Applying Bayes' theorem

\begin{equation}
    P(p \mid X) = \frac{P(X \mid p) \cdot P(p)}{P(X)}
\end{equation}

we find the posterior of the rogue wave probability as:

\begin{equation}
    p \sim \operatorname{Beta}(n^+ + \alpha_0,\; n^- + \beta_0) \label{eq:p-posterior}
\end{equation}

I.e., another Beta distribution (since the chosen Beta prior for $p$ is conjugate to the binomial likelihood of $n^+$).

This posterior is simple to evaluate analytically. In particular, we can use widely available library functions to compute the minimum credible interval (highest posterior credible interval) for $p$. This gives us the possibility to quantify our uncertainty in $p$ based on the number of available samples, expressed as e.g. the 68\% and 95\% credible interval.

To finally investigate the influence of the sea state on the rogue wave probability $p$, we split each sea state parameter into 15 equally sized bins. We assume that, within each bin, $p$ is independently and identically distributed (\emph{iid.}) with a distribution according to \eqref{eq:p-posterior}, and evaluate the mean and credible interval of $p$ independently for each bin. We also exclude bins that contain less than $10$ rogue wave events (i.e., where $n^+ < 10$) to eliminate overly uncertain estimates.

As a result, we can study how $p$ behaves as a function of each sea state parameter, and quantify our uncertainty based on how much data we have in each regime.

We stress however that this uncertainty is based on the assumption that $p$ is iid.\ Beta distributed within each bin, which is clearly not the case if we acknowledge that $p$ depends on more than one parameter. Therefore, these uncertainties can only serve as an indicator whether or not there is enough data to make a statement about this marginalized version of the true, multivariate distribution of $p$. In other words, they indicate how confident we can be in the best estimate of $p$ for this dataset if we can only measure one parameter at a time.

\begin{figure*}
    \centering
    \includegraphics[height=.9\textheight]{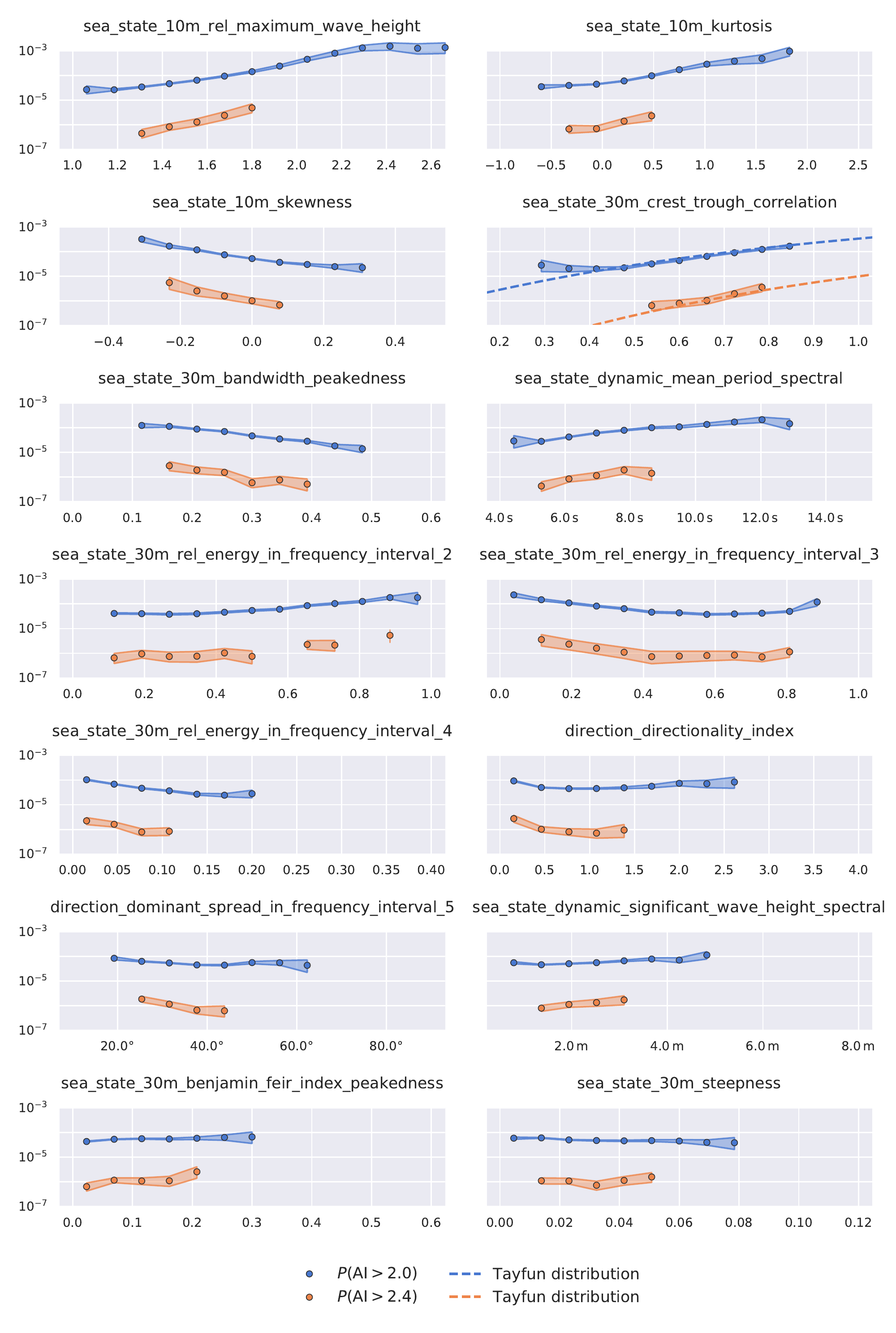}
    \caption{Some sea state parameters are much more informative for rogue wave activity than others. Shown is the dependence of the rogue wave probability on several sea state parameters for $\abni > 2$ and $\abni > 2.4$. Markers represent rogue wave probability posterior mean, shading \SI{95}{\percent} minimum credible interval. Dashed lines indicate the values predicted by the Tayfun wave height distribution \citep{tayfun_wave-height_2007}.} \label{fig:p-rogue-binned}
\end{figure*}

The results of this process show a clear, highly significant dependence of the rogue wave probability on some sea state parameters, and the lack of such a dependence on others (\figref{fig:p-rogue-binned}). In particular, we find that:

\begin{enumerate}
    \item Surface elevation kurtosis, relative maximum wave height, and skewness are the strongest predictors for rogue wave risk. For relative maximum wave height, $P(\abni > 2)$ ranges between $2.9 \times 10^{-5}$ and $1.0 \times 10^{-3}$. So if an up-to-date, in-situ surface elevation time series is available, these parameters are able to quantify rogue wave risk with a factor of about 35 in variation.
    \item Crest-trough correlation and spectral bandwidth (peakedness) are the strongest spectral predictors, with $P(\abni > 2)$ varying between $2.4 \times 10^{-5}$ and $1.4 \times 10^{-4}$ for crest-trough correlation --- i.e.,\ almost 1 order of magnitude in variation from the spectrum alone.
    \item The Tayfun wave height distribution \citep{tayfun_m._aziz_distribution_1990,tayfun_wave-height_2007} seems to be an excellent baseline for rogue wave activity.
    \item There is, at this level of detail, only a minor dependency of rogue wave occurrence on directional spread, Benjamin-Feir index, significant wave height, and steepness.
\end{enumerate}

So, in this first analysis, it seems that bandwidth effects are the dominant modifier of rogue wave risk, while nonlinear effects (at least those governed by steepness and BFI) seem to play a minor corrective role compared to that. However, it is important to keep in mind that we are only looking at one set of stations and only one sea state parameter at a time.

\section{Conclusion} \label{sec:conclusion}

FOWD is a free ocean wave dataset that relates wave point measurements to the conditions in which the wave occurred, and that is optimized for use in data mining and machine learning applications.

In the previous sections, we describe which quantities are included in our wave catalogue FOWD and how they are computed, and which steps we take to ensure quality and reproducibility (\secref{sec:specification}).

We describe the reference implementation and the steps we take to be able to process massive amounts of data at the Terabyte scale (\secref{sec:implementation}).

We summarize the processing of the CDIP buoy data catalogue and analyze the quality of the resulting catalogue (\secref{sec:cdip}). We apply additional filtering to remove problematic measurements. By visual inspection, we find that the resulting dataset is of excellent quality.

Finally, we study the occurrence probability of rogue waves depending on the sea state in an example application, where we have been able to demonstrate that certain parameters are much better predictors than others (\secref{sec:application}). We find that, based on analyzing only one sea state parameter at a time, rogue wave risk can vary by at least 1 order of magnitude. The estimated rogue wave probabilities are consistent with those found in earlier studies based on observations and simulations \citep[e.g.][]{fedele_real_2016,fedele_sinking_2017}.

The strongest parameters in this analysis are surface elevation skewness / kurtosis, and maximum relative wave height of the past record. This is of little surprise when taking into account how many rogue waves occur in rapid succession of each other (\tabref{tab:rogue-stats}), but the importance of kurtosis and skewness could also be evidence for the role of second and third-order (weakly) nonlinear contributions \citep{mori_kurtosis_2006,gemmrich_dynamical_2011,christou_field_2014}. The most important spectral parameters are spectral bandwidth and crest-trough correlation, which is compatible with the finding in \citet{cattrell_can_2018} that spectral bandwidth is important (although we disagree with their conclusion that rogue waves \emph{cannot} be predicted from characteristic parameters).

On the other hand, we were unable to detect any noteworthy dependency of rogue wave risk on directional spread \citep[hypothesized e.g.\ by][]{gramstad_modulational_2018,mcallister_laboratory_2019}, wave steepness (which is evidence against the importance of weakly nonlinear corrections), or Benjamin-Feir index \citep[one of two parameters used by ECMWF's freak wave forecast, see][]{janssen_extension_2009}. This does of course \emph{not} prove that such dependencies do not exist, just that it is not detectable in this limited dataset (of Hawaiian stations) and by univariate analysis (i.e., considering one parameter at a time). A more sophisticated analysis is needed, which is precisely what we want to enable with FOWD.

We believe that this work represents an important motivation and contribution to enable physical insight into ocean waves through sophisticated data mining and big data methods. Downstream studies can either process their own raw data --- due to the flexibility of the FOWD specification and reference implementation --- or make use of the already processed CDIP data.

Extreme probabilistic events such as rogue waves are notoriously difficult to analyze statistically in a robust, meaningful way. By lowering the bar of entry for non-wave experts, we hope to enable new, powerful descriptive and predictive approaches to ocean wave phenomena.

\acknowledgments
Dion Häfner received funding from the Danish Hydrocarbon Research and Technology Centre (DHRTC).

We thank Øyvind Breivik and Mika Malila for their valuable comments during the drafting stage of FOWD.

We thank 3 anonymous reviewers for their constructive and insightful remarks.

Raw data were furnished by the Coastal Data Information Program (CDIP), Integrative Oceanography Division, operated by the Scripps Institution of Oceanography, under the sponsorship of the U.S. Army Corps of Engineers and the California Department of Parks and Recreation.

Computational resources were provided by DC$^3$, the Danish Center for Climate Computing.

%
%
\datastatement

Filtered and unfiltered versions of the FOWD CDIP data are available for download at \url{https://sid.erda.dk/public/archives/969a4d819822c8f0325cb22a18f64eb8/published-archive.html} (to be replaced with DOI before publication).

The exact version of the FOWD reference implementation used throughout this study (v0.5.2) is available at \url{https://doi.org/10.5281/zenodo.4628203}. The current version can be found at \url{https://github.com/dionhaefner/FOWD}.

The scripts used to generate the plots and statistics in this paper are available at \url{https://gist.github.com/dionhaefner/51ef93980a87d6b6bb557599b79582da}.

%






%
%
%

\appendix

See \quantitytable for an exhaustive list of all quantities included in FOWD.

%
%

\begin{table*}
\begin{tabular}{p{.25\textwidth}p{.4\textwidth}p{.045\textwidth}p{.2\textwidth}}
    \appendcaption{A1}{All quantities included in FOWD output files. Quantities marked with ${}^\dagger$ are further explained throughout \subsecref{sec:specification}{sec:quantities}.}
    \label{tab:quantities} \\
    \multicolumn{4}{c}{Station metadata} \\ \doublerule
    Name in output dataset & Description & Unit & Example value \\ \midrule
    \param{meta\_station\_name} & Name of original measurement station & --- & \param{CDIP\_098p1} \\
    \param{meta\_source\_file\_name} & File name of raw input data file & --- & \param{098p1\_d01.nc} \\
    \param{meta\_source\_file\_uuid} & UUID of raw input data file & --- & \param{CC54C8D5\allowbreak-7B1B\allowbreak-4170\allowbreak-9DBA\allowbreak-EBFD91F26F14} \\
    \param{meta\_deploy\_latitude} & Deploy latitude of instrument & $^\circ$N & \param{21.4156} \\
    \param{meta\_deploy\_longitude} & Deploy longitude of instrument & $^\circ$E & \param{-157.678} \\
    \param{meta\_water\_depth} & Water depth at deployment location & \si{\metre} & \param{100.0} \\
    \param{meta\_sampling\_rate} & Measurement sampling frequency in time & \si{\hertz} & \param{1.28} \\
    \param{meta\_frequency\_band\_lower} & Lower limit of frequency band & \si{\hertz} & \param{[0.0, 0.05, 0.1, 0.25, 0.08]} \\
    \param{meta\_frequency\_band\_upper} & Upper limit of frequency band & \si{\hertz} & \param{[0.05, 0.1, 0.25, 1.5, 0.5]} \\
    \bottomrule \\[1em]
    \end{tabular}
\end{table*}
\begin{table*}
\begin{tabular}{p{.25\textwidth}p{.4\textwidth}p{.045\textwidth}p{.2\textwidth}}
    \multicolumn{4}{c}{Wave-specific parameters} \\ \doublerule
    Name in output dataset & Description & Unit & Example value \\ \midrule
    \param{wave\_id\_local} & Incrementing wave ID for given station & --- & \param{11726} \\
    \param{wave\_start\_time} & Wave start time & --- & \param{2000-08-10T\allowbreak12:18:44.220000000} \\
    \param{wave\_end\_time} & Wave end time & --- & \param{2000-08-10T\allowbreak12:18:50.470000000} \\
    \param{wave\_zero\_crossing\_period} & Wave zero-crossing period relative to 30m sea surface elevation & \si{\second} & \param{5.644304276} \\
    \param{wave\_zero\_crossing\_wavelength}${}^\dagger$ & Wave zero-crossing wavelength relative to 30m sea surface elevation & \si{\metre} & \param{49.74048} \\
    \param{wave\_raw\_elevation} & Raw surface elevation relative to 30m sea surface elevation & \si{\metre} & \param{[0.200261, 0.889527, 0.509184, -0.550564, -0.690152, -0.270083, -0.200052]} \\
    \param{wave\_crest\_height} & Wave crest height relative to 30m sea surface elevation & \si{\metre} & \param{0.889527} \\
    \param{wave\_trough\_depth} & Wave trough depth relative to 30m sea surface elevation & \si{\metre} & \param{-0.690152} \\
    \param{wave\_height} & Absolute wave height relative to 30m sea surface elevation & \si{\metre} & \param{1.579679} \\
    \param{wave\_ursell\_number} & Ursell number & 1 & \param{0.003908} \\
    \param{wave\_maximum\_elevation\_slope} & Maximum slope of surface elevation in time & \si{\metre\per\second} & \param{0.921658} \\
    \bottomrule \\[1em]
    \end{tabular}
\end{table*}
\begin{table*}
\begin{tabular}{p{.25\textwidth}p{.4\textwidth}p{.045\textwidth}p{.2\textwidth}}
    \multicolumn{4}{c}{Aggregated sea state parameters} \\ \doublerule
    Name in output dataset & Description & Unit & Example value \\ \midrule
    \param{sea\_state\_30m\_start\_time} & Sea state aggregation start time & --- & \param{2000-08-10T\allowbreak11:48:45.000999936} \\
    \param{sea\_state\_30m\_end\_time} & Sea state aggregation end time & --- & \param{2000-08-10T\allowbreak12:18:43.438000000} \\
    \param{sea\_state\_30m\_significant\_wave\_height\_spectral}${}^\dagger$ & Significant wave height estimated from wave spectrum (Hm0) & \si{\metre} & \param{1.798395} \\
    \param{sea\_state\_30m\_significant\_wave\_height\_direct} & Significant wave height estimated from wave history (H1/3) & \si{\metre} & \param{1.648174} \\
    \param{sea\_state\_30m\_maximum\_wave\_height} & Maximum wave height estimated from wave history & \si{\metre} & \param{3.18891} \\
    \param{sea\_state\_30m\_rel\_maximum\_wave\_height} & Maximum wave height estimated from wave history relative to spectral significant wave height & 1 & \param{1.773198} \\
    \param{sea\_state\_30m\_mean\_period\_direct} & Mean zero-crossing period estimated from wave history & \si{\second} & \param{5.133130549} \\
    \param{sea\_state\_30m\_mean\_period\_spectral} & Mean zero-crossing period estimated from wave spectrum & \si{\second} & \param{5.034029007} \\
    \param{sea\_state\_30m\_skewness} & Skewness of sea surface elevation & 1 & \param{0.010083} \\
    \param{sea\_state\_30m\_kurtosis} & Excess kurtosis of sea surface elevation & 1 & \param{-0.076898} \\
    \param{sea\_state\_30m\_valid\_data\_ratio} & Ratio of valid measurements to all measurements & 1 & \param{1.0} \\
    \param{sea\_state\_30m\_peak\_wave\_period}${}^\dagger$ & Dominant wave period & \si{\second} & \param{6.841089249} \\
    \param{sea\_state\_30m\_peak\_wavelength}${}^\dagger$ & Dominant wavelength & \si{\metre} & \param{73.07008} \\
    \param{sea\_state\_30m\_steepness}${}^\dagger$ & Dominant wave steepness & 1 & \param{0.054674} \\
    \param{sea\_state\_30m\_bandwidth\_peakedness}${}^\dagger$ & Spectral bandwidth estimated through spectral peakedness (quality factor) & 1 & \param{0.312186} \\
    \param{sea\_state\_30m\_bandwidth\_narrowness}${}^\dagger$ & Spectral bandwidth estimated through spectral narrowness & 1 & \param{0.43569} \\
    \param{sea\_state\_30m\_benjamin\_feir\_index\_peakedness}${}^\dagger$ & Benjamin-Feir index estimated through steepness and peakedness & 1 & \param{0.164307} \\
    \param{sea\_state\_30m\_benjamin\_feir\_index\_narrowness}${}^\dagger$ & Benjamin-Feir index estimated through steepness and narrowness & 1 & \param{0.117731} \\
    \param{sea\_state\_30m\_crest\_trough\_correlation} & Crest-trough correlation parameter (r) estimated from spectral density & 1 & \param{0.608416} \\
    \param{sea\_state\_30m\_energy\_in\_frequency\_interval}${}^\dagger$ & Total energy density contained in frequency band & \si{\joule\per\metre\squared} & \param{[1.935885, 106.74948, 1620.2413, 301.649, 1926.3574]} \\
    \param{sea\_state\_30m\_rel\_energy\_in\_frequency\_interval}${}^\dagger$ & Relative energy contained in frequency band & 1 & \param{[0.000953, 0.052571, 0.797922, 0.148553, 0.948675]} \\[2em]
    \multicolumn{4}{c}{\textit{Repeated analogously for 10 minute (\texttt{\_10m\_}) and dynamic (\texttt{\_dynamic\_}) window sizes}} \\
    \bottomrule \\[1em]
    \end{tabular}
\end{table*}
\begin{table*}
\begin{tabular}{p{.25\textwidth}p{.4\textwidth}p{.045\textwidth}p{.2\textwidth}}
    \multicolumn{4}{c}{Directional sea state parameters} \\ \doublerule
    Name in output dataset & Description & Unit & Example value \\ \midrule
    \param{direction\_sampling\_time} & Time at which directional quantities are sampled & --- & \param{2000-08-10T\allowbreak12:11:52.000000000} \\
    \param{direction\_dominant\_spread\_in\_frequency\_interval}${}^\dagger$ & Dominant directional spread in frequency band & $^\circ$ & \param{[57.965824, 38.118546, 31.54562, 39.30281, 33.07898]} \\
    \param{direction\_dominant\_direction\_in\_frequency\_interval}${}^\dagger$ & Dominant wave direction in frequency band & $^\circ$ & \param{[83.074, 136.02432, 74.00862, 77.26602, 74.89502]} \\
    \param{direction\_peak\_wave\_direction} & Peak wave direction relative to normal-north & $^\circ$ & \param{70.46875} \\
    \param{direction\_directionality\_index}${}^\dagger$ & Directionality index R (squared ratio of directional spread and spectral bandwidth) & 1 & \param{0.924404} \\
    \bottomrule
\end{tabular}
\end{table*}



\bibliographystyle{ametsoc2014}
\bibliography{references}

\begin{thebibliography}{41}
\providecommand{\natexlab}[1]{#1}
\providecommand{\url}[1]{\texttt{#1}}
\renewcommand{\UrlFont}{\rmfamily}
\providecommand{\urlprefix}{URL }
\expandafter\ifx\csname urlstyle\endcsname\relax
  \providecommand{\doi}[1]{doi:\discretionary{}{}{}#1}\else
  \providecommand{\doi}{doi:\discretionary{}{}{}\begingroup
  \urlstyle{rm}\Url}\fi
\providecommand{\eprint}[2][]{\url{#2}}

\bibitem[{Adcock and Taylor(2014)Adcock, and Taylor}]{adcock_physics_2014}
Adcock, T. A.~A., and P.~H. Taylor, 2014: The physics of anomalous
  (‘rogue’) ocean waves. \textit{Reports on Progress in Physics},
  \textbf{77~(10)}, 105\,901, \doi{10.1088/0034-4885/77/10/105901}.

\bibitem[{Barbariol et~al.(2019)Barbariol, Bidlot, Cavaleri, Sclavo, Thomson,,
  and Benetazzo}]{barbariol_maximum_2019}
Barbariol, F., J.-R. Bidlot, L.~Cavaleri, M.~Sclavo, J.~Thomson, and
  A.~Benetazzo, 2019: Maximum wave heights from global model reanalysis.
  \textit{Progress in Oceanography}, \textbf{175}, 139--160,
  \doi{10.1016/j.pocean.2019.03.009},
  \urlprefix\url{http://www.sciencedirect.com/science/article/pii/S0079661118302763}.

\bibitem[{Behrens et~al.(2019)Behrens, Thomas, Terrill,, and
  Jensen}]{behrens_cdip_2019}
Behrens, J., J.~Thomas, E.~Terrill, and R.~Jensen, 2019: {CDIP}: {Maintaining}
  a {Robust} and {Reliable} {Ocean} {Observing} {Buoy} {Network}. \textit{2019
  {IEEE}/{OES} {Twelfth} {Current}, {Waves} and {Turbulence} {Measurement}
  ({CWTM})}, 1--5, \doi{10.1109/CWTM43797.2019.8955166}.

\bibitem[{Benetazzo et~al.(2015)Benetazzo, Barbariol, Bergamasco, Torsello,
  Carniel,, and Sclavo}]{benetazzo_observation_2015}
Benetazzo, A., F.~Barbariol, F.~Bergamasco, A.~Torsello, S.~Carniel, and
  M.~Sclavo, 2015: Observation of {Extreme} {Sea} {Waves} in a {Space}–{Time}
  {Ensemble}. \textit{Journal of Physical Oceanography}, \textbf{45~(9)},
  2261--2275, \doi{10.1175/JPO-D-15-0017.1},
  \urlprefix\url{https://journals.ametsoc.org/doi/10.1175/JPO-D-15-0017.1}.

\bibitem[{Boccotti(2000)}]{boccotti_wave_2000}
Boccotti, P., 2000: \textit{Wave {Mechanics} for {Ocean} {Engineering}}.
  Elsevier, google-Books-ID: 1319kgDa8GUC.

\bibitem[{Casas‐Prat and Holthuijsen(2010)Casas‐Prat, and
  Holthuijsen}]{casasprat_short-term_2010}
Casas‐Prat, M., and L.~H. Holthuijsen, 2010: Short-term statistics of waves
  observed in deep water. \textit{Journal of Geophysical Research: Oceans},
  \textbf{115~(C9)}, \doi{10.1029/2009JC005742}.

\bibitem[{Cattrell et~al.(2018)Cattrell, Srokosz, Moat,, and
  Marsh}]{cattrell_can_2018}
Cattrell, A.~D., M.~Srokosz, B.~I. Moat, and R.~Marsh, 2018: Can {Rogue}
  {Waves} {Be} {Predicted} {Using} {Characteristic} {Wave} {Parameters}?
  \textit{Journal of Geophysical Research: Oceans}, \textbf{123~(8)},
  5624--5636, \doi{10.1029/2018JC013958}.

\bibitem[{Christou and Ewans(2014)Christou, and Ewans}]{christou_field_2014}
Christou, M., and K.~Ewans, 2014: Field {Measurements} of {Rogue} {Water}
  {Waves}. \textit{Journal of Physical Oceanography}, \textbf{44~(9)},
  2317--2335, \doi{10.1175/JPO-D-13-0199.1}.

\bibitem[{Dudley et~al.(2019)Dudley, Genty, Mussot, Chabchoub,, and
  Dias}]{dudley_rogue_2019}
Dudley, J.~M., G.~Genty, A.~Mussot, A.~Chabchoub, and F.~Dias, 2019: Rogue
  waves and analogies in optics and oceanography. \textit{Nature Reviews
  Physics}, \textbf{1~(11)}, 675--689, \doi{10.1038/s42254-019-0100-0}, number:
  11 Publisher: Nature Publishing Group.

\bibitem[{Dysthe et~al.(2008)Dysthe, Krogstad,, and
  Müller}]{dysthe_oceanic_2008}
Dysthe, K., H.~E. Krogstad, and P.~Müller, 2008: Oceanic {Rogue} {Waves}.
  \textit{Annual Review of Fluid Mechanics}, \textbf{40~(1)}, 287--310,
  \doi{10.1146/annurev.fluid.40.111406.102203}.

\bibitem[{Fedele(2015)}]{fedele_kurtosis_2015}
Fedele, F., 2015: On the kurtosis of deep-water gravity waves. \textit{Journal
  of Fluid Mechanics}, \textbf{782}, 25--36, \doi{10.1017/jfm.2015.538},
  \urlprefix\url{https://www.cambridge.org/core/journals/journal-of-fluid-mechanics/article/on-the-kurtosis-of-deepwater-gravity-waves/7DE34CF393A3E9A928208A48ACA2C05D},
  publisher: Cambridge University Press.

\bibitem[{Fedele et~al.(2016)Fedele, Brennan, Ponce~de León, Dudley,, and
  Dias}]{fedele_real_2016}
Fedele, F., J.~Brennan, S.~Ponce~de León, J.~Dudley, and F.~Dias, 2016: Real
  world ocean rogue waves explained without the modulational instability.
  \textit{Scientific Reports}, \textbf{6}, 27\,715, \doi{10.1038/srep27715}.

\bibitem[{Fedele et~al.(2019)Fedele, Herterich, Tayfun,, and
  Dias}]{fedele_large_2019}
Fedele, F., J.~Herterich, A.~Tayfun, and F.~Dias, 2019: Large nearshore storm
  waves off the {Irish} coast. \textit{Scientific Reports}, \textbf{9~(1)},
  15\,406, \doi{10.1038/s41598-019-51706-8},
  \urlprefix\url{https://www.nature.com/articles/s41598-019-51706-8}, number: 1
  Publisher: Nature Publishing Group.

\bibitem[{Fedele et~al.(2017)Fedele, Lugni,, and Chawla}]{fedele_sinking_2017}
Fedele, F., C.~Lugni, and A.~Chawla, 2017: The sinking of the {El} {Faro}:
  predicting real world rogue waves during {Hurricane} {Joaquin}.
  \textit{Scientific Reports}, \textbf{7~(1)}, 1--15,
  \doi{10.1038/s41598-017-11505-5},
  \urlprefix\url{https://www.nature.com/articles/s41598-017-11505-5}.

\bibitem[{Fenton(1988)}]{fenton_numerical_1988}
Fenton, J.~D., 1988: The numerical solution of steady water wave problems.
  \textit{Computers \& Geosciences}, \textbf{14~(3)}, 357--368,
  \doi{10.1016/0098-3004(88)90066-0}.

\bibitem[{Gemmrich and Garrett(2011)Gemmrich, and
  Garrett}]{gemmrich_dynamical_2011}
Gemmrich, J., and C.~Garrett, 2011: Dynamical and statistical explanations of
  observed occurrence rates of rogue waves. \textit{Natural Hazards and Earth
  System Science}, \textbf{11~(5)}, 1437--1446,
  \doi{10.5194/nhess-11-1437-2011}.

\bibitem[{Gemmrich et~al.(2016)Gemmrich, Thomson, Rogers, Pleskachevsky,, and
  Lehner}]{gemmrich_spatial_2016}
Gemmrich, J., J.~Thomson, W.~E. Rogers, A.~Pleskachevsky, and S.~Lehner, 2016:
  Spatial characteristics of ocean surface waves. \textit{Ocean Dynamics},
  \textbf{66~(8)}, 1025--1035, \doi{10.1007/s10236-016-0967-6}.

\bibitem[{Gramstad et~al.(2018)Gramstad, Bitner-Gregersen, Trulsen,, and
  Nieto~Borge}]{gramstad_modulational_2018}
Gramstad, O., E.~Bitner-Gregersen, K.~Trulsen, and J.~C. Nieto~Borge, 2018:
  Modulational {Instability} and {Rogue} {Waves} in {Crossing} {Sea} {States}.
  \textit{Journal of Physical Oceanography}, \textbf{48~(6)}, 1317--1331,
  \doi{10.1175/JPO-D-18-0006.1}.

\bibitem[{Haver(2004)}]{haver2004possible}
Haver, S., 2004: A possible freak wave event measured at the {D}raupner
  {J}acket {J}anuary 1 1995.
  \urlprefix\url{http://www.ifremer.fr/web-com/stw2004/rw/fullpapers/walk_on_haver.pdf}.

\bibitem[{Holthuijsen(2010)}]{holthuijsen_waves_2010}
Holthuijsen, L.~H., 2010: \textit{Waves in {Oceanic} and {Coastal} {Waters}}.
  Cambridge University Press, google-Books-ID: 7tFUL2blHdoC.

\bibitem[{Huber(2011)}]{Huber2011}
Huber, P.~J., 2011: \textit{Robust Statistics}, 1248--1251. Springer Berlin
  Heidelberg, Berlin, Heidelberg, \doi{10.1007/978-3-642-04898-2_594},
  \urlprefix\url{https://doi.org/10.1007/978-3-642-04898-2_594}.

\bibitem[{Janssen and Bidlot(2009)Janssen, and Bidlot}]{janssen_extension_2009}
Janssen, P., and J.-R. Bidlot, 2009: On the extension of the freak wave warning
  system and its verification. \doi{10.21957/uf1sybog}.

\bibitem[{Janssen(2003)}]{janssen_nonlinear_2003}
Janssen, P. A. E.~M., 2003: Nonlinear {Four}-{Wave} {Interactions} and {Freak}
  {Waves}. \textit{Journal of Physical Oceanography}, \textbf{33~(4)},
  863--884, \doi{10.1175/1520-0485(2003)33<863:NFIAFW>2.0.CO;2},
  \urlprefix\url{https://journals.ametsoc.org/jpo/article/33/4/863/68537/Nonlinear-Four-Wave-Interactions-and-Freak-Waves},
  publisher: American Meteorological Society.

\bibitem[{Karmpadakis et~al.(2020)Karmpadakis, Swan,, and
  Christou}]{karmpadakis_assessment_2020}
Karmpadakis, I., C.~Swan, and M.~Christou, 2020: Assessment of wave height
  distributions using an extensive field database. \textit{Coastal
  Engineering}, 103630, \doi{10.1016/j.coastaleng.2019.103630}.

\bibitem[{Kharif and Pelinovsky(2003)Kharif, and
  Pelinovsky}]{kharif_physical_2003}
Kharif, C., and E.~Pelinovsky, 2003: Physical mechanisms of the rogue wave
  phenomenon. \textit{European Journal of Mechanics - B/Fluids},
  \textbf{22~(6)}, 603--634, \doi{10.1016/j.euromechflu.2003.09.002}.

\bibitem[{Longuet-Higgins(1952)}]{longuet1952statisticaldistribution}
Longuet-Higgins, M.~S., 1952: On the statistical distribution of the height of
  sea waves. \textit{JMR}, \textbf{11}, 245--266.

\bibitem[{McAllister et~al.(2019)McAllister, Draycott, Adcock, Taylor,, and
  Bremer}]{mcallister_laboratory_2019}
McAllister, M.~L., S.~Draycott, T.~a.~A. Adcock, P.~H. Taylor, and T.~S. v.~d.
  Bremer, 2019: Laboratory recreation of the {Draupner} wave and the role of
  breaking in crossing seas. \textit{Journal of Fluid Mechanics}, \textbf{860},
  767--786, \doi{10.1017/jfm.2018.886}.

\bibitem[{McAllister and van~den Bremer(2019{\natexlab{a}})McAllister, and
  van~den Bremer}]{mcallister_experimental_2019}
McAllister, M.~L., and T.~S. van~den Bremer, 2019{\natexlab{a}}: Experimental
  {Study} of the {Statistical} {Properties} of {Directionally} {Spread} {Ocean}
  {Waves} {Measured} by {Buoys}. \textit{Journal of Physical Oceanography},
  \textbf{50~(2)}, 399--414, \doi{10.1175/JPO-D-19-0228.1}, publisher: American
  Meteorological Society.

\bibitem[{McAllister and van~den Bremer(2019{\natexlab{b}})McAllister, and
  van~den Bremer}]{mcallister_lagrangian_2019}
McAllister, M.~L., and T.~S. van~den Bremer, 2019{\natexlab{b}}: Lagrangian
  {Measurement} of {Steep} {Directionally} {Spread} {Ocean} {Waves}:
  {Second}-{Order} {Motion} of a {Wave}-{Following} {Measurement} {Buoy}.
  \textit{Journal of Physical Oceanography}, \textbf{49~(12)}, 3087--3108,
  \doi{10.1175/JPO-D-19-0170.1}, publisher: American Meteorological Society.

\bibitem[{Mori and Janssen(2006)Mori, and Janssen}]{mori_kurtosis_2006}
Mori, N., and P.~A. E.~M. Janssen, 2006: On {Kurtosis} and {Occurrence}
  {Probability} of {Freak} {Waves}. \textit{Journal of Physical Oceanography},
  \textbf{36~(7)}, 1471--1483, \doi{10.1175/JPO2922.1}, publisher: American
  Meteorological Society.

\bibitem[{Ochi and Hubble(1977)Ochi, and Hubble}]{ochi1977six}
Ochi, M.~K., and E.~N. Hubble, 1977: Six-parameter wave spectra.
  \textit{Coastal Engineering 1976}, 301--328.

\bibitem[{Portilla‐Yandún(2018)}]{portillayandun_global_2018}
Portilla‐Yandún, J., 2018: The {Global} {Signature} of {Ocean} {Wave}
  {Spectra}. \textit{Geophysical Research Letters}, \textbf{45~(1)}, 267--276,
  \doi{https://doi.org/10.1002/2017GL076431},
  \urlprefix\url{https://agupubs.onlinelibrary.wiley.com/doi/abs/10.1002/2017GL076431},
  \_eprint:
  https://agupubs.onlinelibrary.wiley.com/doi/pdf/10.1002/2017GL076431.

\bibitem[{Portilla‐Yandún et~al.(2016)Portilla‐Yandún, Salazar,, and
  Cavaleri}]{portillayandun_climate_2016}
Portilla‐Yandún, J., A.~Salazar, and L.~Cavaleri, 2016: Climate patterns
  derived from ocean wave spectra. \textit{Geophysical Research Letters},
  \textbf{43~(22)}, 11,736--11,743, \doi{https://doi.org/10.1002/2016GL071419},
  \urlprefix\url{https://agupubs.onlinelibrary.wiley.com/doi/abs/10.1002/2016GL071419},
  \_eprint:
  https://agupubs.onlinelibrary.wiley.com/doi/pdf/10.1002/2016GL071419.

\bibitem[{Serio et~al.(2005)Serio, Onorato, R~a Osborne,, and
  Janssen}]{serio_computation_2005}
Serio, M., M.~Onorato, A.~R~a Osborne, and P.~Janssen, 2005: On the computation
  of the {Benjamin}-{Feir} {Index}. \textit{Nuovo Cimento della Societa
  Italiana di Fisica C}, \textbf{28}, 893--903,
  \doi{10.1393/ncc/i2005-10134-1}.

\bibitem[{Slunyaev et~al.(2011)Slunyaev, Didenkulova,, and
  Pelinovsky}]{slunyaev_rogue_2011}
Slunyaev, A., I.~Didenkulova, and E.~Pelinovsky, 2011: Rogue waters.
  \textit{Contemporary Physics}, \textbf{52~(6)}, 571--590,
  \doi{10.1080/00107514.2011.613256}.

\bibitem[{Tayfun(1990)}]{tayfun_m._aziz_distribution_1990}
Tayfun, M.~A., 1990: Distribution of {Large} {Wave} {Heights}. \textit{Journal
  of Waterway, Port, Coastal, and Ocean Engineering}, \textbf{116~(6)},
  686--707, \doi{10.1061/(ASCE)0733-950X(1990)116:6(686)}.

\bibitem[{Tayfun and Fedele(2007)Tayfun, and Fedele}]{tayfun_wave-height_2007}
Tayfun, M.~A., and F.~Fedele, 2007: Wave-height distributions and nonlinear
  effects. \textit{Ocean Engineering}, \textbf{34~(11)}, 1631--1649,
  \doi{10.1016/j.oceaneng.2006.11.006}.

\bibitem[{Toffoli et~al.(2010)Toffoli, Gramstad, Trulsen, Monbaliu,
  Bitner-Gregersen,, and Onorato}]{toffoli_evolution_2010}
Toffoli, A., O.~Gramstad, K.~Trulsen, J.~Monbaliu, E.~Bitner-Gregersen, and
  M.~Onorato, 2010: Evolution of weakly nonlinear random directional waves:
  laboratory experiments and numerical simulations. \textit{Journal of Fluid
  Mechanics}, \textbf{664}, 313--336, \doi{10.1017/S002211201000385X},
  publisher: Cambridge University Press.

\bibitem[{Welch(1967)}]{welch_use_1967}
Welch, P., 1967: The use of fast {Fourier} transform for the estimation of
  power spectra: {A} method based on time averaging over short, modified
  periodograms. \textit{IEEE Transactions on Audio and Electroacoustics},
  \textbf{15~(2)}, 70--73, \doi{10.1109/TAU.1967.1161901}, conference Name:
  IEEE Transactions on Audio and Electroacoustics.

\bibitem[{Xiao et~al.(2013)Xiao, Liu, Wu,, and Yue}]{xiao_rogue_2013}
Xiao, W., Y.~Liu, G.~Wu, and D.~K.~P. Yue, 2013: Rogue wave occurrence and
  dynamics by direct simulations of nonlinear wave-field evolution.
  \textit{Journal of Fluid Mechanics}, \textbf{720}, 357--392,
  \doi{10.1017/jfm.2013.37}, publisher: Cambridge University Press.

\bibitem[{Young(1995)}]{young_determination_1995}
Young, I.~R., 1995: The determination of confidence limits associated with
  estimates of the spectral peak frequency. \textit{Ocean Engineering},
  \textbf{22~(7)}, 669--686, \doi{10.1016/0029-8018(95)00002-3}.

\end{thebibliography}

%

%

\end{document}